\documentclass[aps, prb, twocolumn,floatfix, superscriptaddress]{revtex4-2}
    \usepackage{graphicx} 
    \usepackage{float}
    \usepackage[utf8]{inputenc}
    \usepackage[T1]{fontenc}
    \usepackage[english]{babel}
    \usepackage{amsmath}
    \usepackage{amssymb}
    \usepackage{xcolor}
    \usepackage{braket}
    \usepackage{soul}
    \usepackage{bm}
    \usepackage{physics}
    \usepackage[colorlinks=true, linkcolor=blue,
    citecolor=red, 
    urlcolor=blue, hypertexnames=true]{hyperref}
    \definecolor{darkgreen}{rgb}{0,0.5,0}

    \usepackage{multibib}
    \begin{document}
    ﻿
    \title{Generating pairwise entanglement in periodically driven quantum spin chains with stochastic resetting}
    ﻿
    ﻿
    \author{Sinchan Ghosh}
    \email{sinchanghosh008@gmail.com}
    \affiliation{School of Physical Sciences, Indian Association for the Cultivation of Science,
    Jadavpur 700032, India}
    ﻿
    ﻿
    \author{Manas Kulkarni}
    \email{manas.kulkarni@icts.res.in}
    \affiliation{International Centre for Theoretical Sciences, Tata Institute of Fundamental Research,
    Bengaluru 560089, India}
    ﻿
    ﻿
    \author{K. Sengupta}
    \email{ksengupta1@gmail.com}
    \affiliation{School of Physical Sciences, Indian Association for the Cultivation of Science,
    Jadavpur 700032, India}
    ﻿
    ﻿
    \author{Satya N. Majumdar}
    \email{satyanarayan.majumdar@cnrs.fr }
    \affiliation{LPTMS, CNRS, Univ. Paris-Sud, Université Paris-Saclay,
    91405 Orsay, France}
    ﻿
    ﻿
    ﻿
    \date{\today}
    ﻿
    \begin{abstract}
    We show that stochastic resetting may lead to finite entanglement between individual, spatially separated spins (pairwise entanglement) in the steady state of the spin chains driven periodically with frequency $\omega_D$. We find the presence of a critical resetting rate $r_c$ below which the steady state pairwise entanglement, measured via concurrence $C$, vanishes. We also identify an optimal resetting rate $r_m$ at which $C$ becomes maximum. These critical and optimal rates exhibit a non-monotonic dependence on $\omega_D$. Our analysis demonstrates the existence of special drive frequencies at which $r_c$ vanishes and $r_m$ attains minima. We compute $C$ in the presence of stochastic resetting using exact diagonalization for both the integrable XY model and non-integrable Rydberg spin chains, which demonstrate these features. Our numerical results match perturbative analytical expressions for the special drive frequencies in the large drive amplitude regime. 
    \end{abstract}
    \maketitle
    
    Entanglement, characterized by von-Neumann or Renyi entropies for pure states, provides a way to measure correlation between macroscopic subsystems of a quantum many-body system \cite{Bennett1996,BZ2006, VV_RMP,L2016}. This measure captures global aspects of entanglement in a many-body system irrespective of the subsystem partition~\cite{Meyer2002}. In contrast, entanglement between two individual, spatially separated, spins or particles in a many-body quantum system is characterized by their concurrence $C$ (or equivalently, negativity) \cite{WW_PRL,HW_PRL,H4_PRL,W_inf,M_PRL,GW_PRA,KZ_PRA,Peres_PRL}. This distinct entanglement measure is necessary because the reduced density matrix of two spins, in general, represents a mixed state obtained after tracing out the rest of the system. The von-Neumann entropy ($S$) and the concurrence ($C$) may exhibit drastically different behaviors depending on the quantum state of the underlying many-body system. For example, the state of a quantum Ising chain after a linear ramp may feature transition from zero to  near-zero concurrence or negativity while maintaining a finite entanglement ($S$) among the subsystems~\cite{dsen1,ramp2,Arul2005}.
    
    Recently, there has been a surge of interest in studying properties of quantum dynamics of periodically driven systems \cite{rev1,rev2,rev3,rev4,rev5, rev6,rev7,rev8,rev9,rev10,rev11,rev12,rev13,rev14,rev15,rev16,rev17,EA_RMP,oka_annurev}. These drives are characterized by a frequency $\omega_D$ and a drive amplitude $\lambda_0$. For large $\lambda_0$ or $\omega_D$, such driven systems exhibit a long prethermal regime before reaching a steady state~\cite{Pth_1,Pth_2,Pth_3,Pth_4,Pth_5}. Such steady states are described by infinite-temperature thermal ensembles for ergodic and generalized Gibbs ensembles for integrable systems \cite{rev2,rev6}. In the prethermal regime, they are known to exhibit emergent symmetries leading to several interesting phenomena \cite{rev17}. Usually, the driven state in such prethermal regime (even the true steady state for integrable models) has a finite $S$ obeying volume law. However, the two-spin entanglement, measured via concurrence $C$, has not been studied so far in such systems.
    
    \begin{figure}
    \includegraphics[width=\columnwidth]{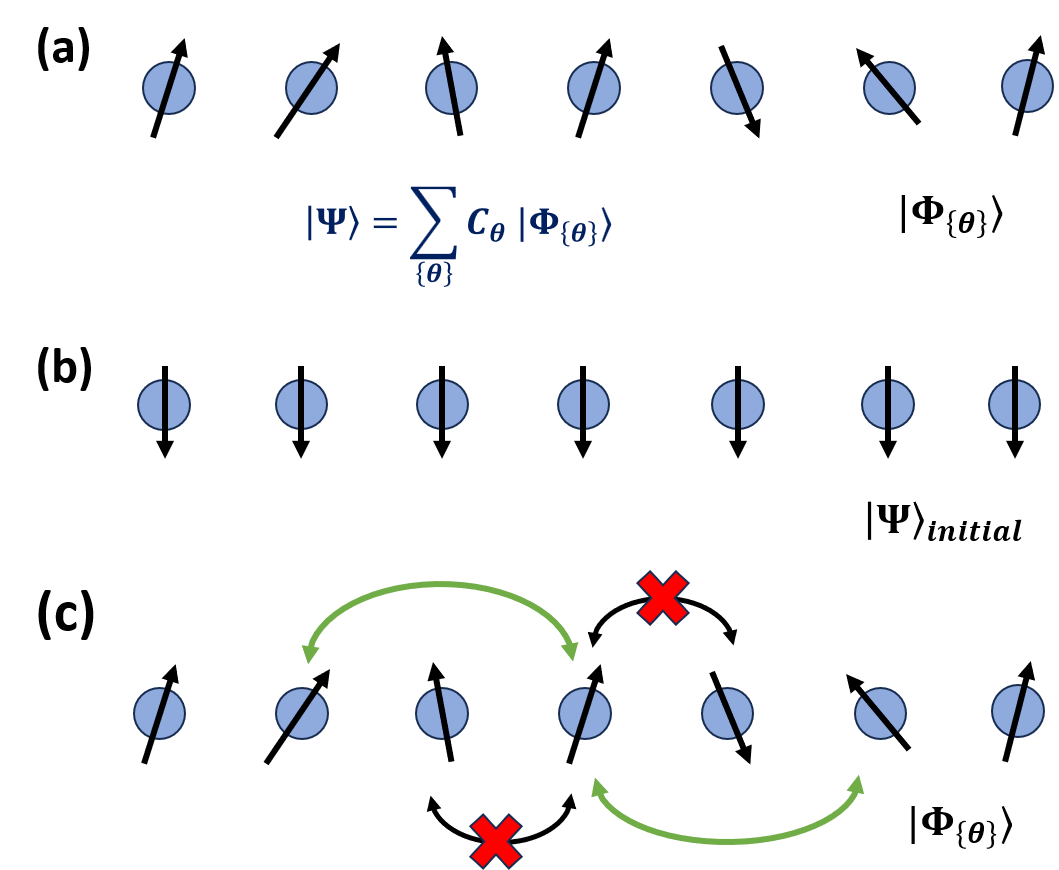}
    \caption{ (a) Schematic representation of the state $|\psi(t)\rangle$ of a spin chain after $m$ drive cycles. The state typically involves superposition of several Fock states $|\Phi_{\{\theta\}}\rangle$. (b) The state of the spin chain right after a reset. (c) Schematic representation of entanglement between two spins in the chain separated by one site ($l=2)$. For the chains considered in this work, the nearest neighbor ($l=1$) pairwise entanglement is either zero due to symmetry of the model (XY chain) or constraint (Rydberg chain). See text for details.  \label{fig0}}
    \end{figure}
    
    Stochastic resetting, studied in many classical systems since its introduction~\cite{Satya_PRL,Evans_2011}, has found a multitude of applications across disciplines~\cite{Evans_2020,GJ_review}. Stochastic resetting simply means interruption and restart of the natural dynamics of a system at random Poisson distributed times where the natural dynamics can be classical or quantum, deterministic or stochastic. For example, in quantum systems, stochastic resetting involves interruption of the unitary deterministic dynamics at random times. There has been a recent surge of interest in quantum systems driven by stochastic resetting ~\cite{Mukherjee_2018,Rose2018spectral,Perfetto2021,Magoni_2022,Miguel_PRX,KM_PRA_23,Kulkarni_2023,Barkai_PRL,Puente_quantum,gotta2026,YWTB25,king2026_arxiv, jafari2026,Wald2025, carollo2026}.
    
    The steady state of a periodically driven system mentioned earlier becomes athermal in the presence of stochastic resetting ~\cite{Mukherjee_2018,Rose2018spectral,Nagar_2023}. For such a protocol, the state $|\psi(t)\rangle$, schematically represented in Fig.~\ref{fig0}(a), is stochastically reset to a predetermined initial state $|\psi(0)\rangle$ [Fig.~\ref{fig0}(b)]. The times for such a reset are typically chosen from a Poissonian distribution characterized by a rate $r$. The dynamics and the steady state of such a system can be significantly different from its counterpart without resetting. For example, such steady states can not be described by the diagonal ensembles (DE) as predicted by the eigenstate thermalization hypothesis (ETH) \cite{MS1,JD1,rev6}. The behavior of entanglement between individual spins, as quantified by concurrence $C$, has not been studied for such steady states of a many-body quantum system. 
    
    In this work, we study entanglement between individual spins in spin chains driven periodically with frequency $\omega_D$ and amplitude $\lambda_0$, and subjected to stochastic resetting characterized by a rate $r$. We show that, in the absence of resetting and for generic drive frequencies, the steady state of such a driven system does not host pairwise entanglement between spins and has $C=0$. In contrast, in the presence of stochastic resetting, $C$ is finite and exhibits non-monotonic behavior as a function of $r$. It becomes non-zero beyond a critical resetting rate $r_c$ and reaches a maximum value at an optimal rate $r_m$. We find that $r_c$ and $r_m$ show non-monotonic dependence on $\lambda_0$ and $\omega_D$. Our analysis indicates existence of special drive frequencies $\omega_n^*= \lambda_0/(n \hbar)$ (where $\hbar$ is the Planck's constant and $n\in \mathbb{Z}$) around which $r_c=0$ and $r_m$ attains a minimum.  Ref.~\cite{KM_PRA_23} considered a non-driven model of two interacting spins. Our work is a significant step forward and a generalization to Floquet many-body systems where drive frequency tuning is a crucial aspect. Studies in Ref.~\cite{KM_PRA_23} indicate existence of optimal reset rate in concurrence but do not predict critical reset rate because entanglement between two spins in an interacting two-body system never vanishes in the non-resetting steady state. Our analytical results are supported by numerical exact diagonalization (ED) studies carried on integrable ${\rm XY}$ and non-integrable Rydberg spin chains \cite{rev1,rev3, exp1,exp2,exp3,exp4}. These results demonstrate that stochastic resetting may provide a way of generating finite pairwise entanglement in the steady states of such driven quantum systems. Experimental feasibility of these protocols have been discussed towards the end of the paper.
    
    {\it Driven spin models}: We consider two driven spin-half chains for which the spin on site $j$ are represented by Pauli matrices $\sigma_j^{\alpha}$, where $\alpha=x,y,z$. The Hamiltonians for these chains are given by
    \begin{eqnarray} 
    \label{xyham} 
    H_1(t) &=& -\frac{J}{4} \sum_{\langle i j\rangle } ((1+\kappa)\sigma_i^x \sigma_j^x +(1-\kappa) \sigma_i^y \sigma_j^y) \nonumber\\
    && -\lambda(t) \sum_j \sigma_j^z,\label{xyham} \\
    H_2(t) &=& \sum_j ( w \tilde \sigma_j^x -\lambda(t) \tilde \sigma_j^z), \quad \tilde \sigma_j^x = P_{j-1} \sigma_j^x P_{j+1}. \label{pxpham} 
    \end{eqnarray} 
    Here $H_1(t)$ denotes the Hamiltonian of a driven integrable ${\rm XY}$ chain, while $H_2(t)$ denotes that for the PXP chain. The latter constitutes a non-integrable model and represents the effective low-energy Hamiltonian for the ultracold Rydberg atoms in an optical lattice \cite{exp1,exp2,exp3,exp4}. In Eq.~\eqref{xyham}, $\langle ij\rangle$ indicates that sites $i$ and $j$ are nearest neighbors, $J>0$ is the amplitude of nearest-neighbor interaction between spins for the ${\rm XY}$ chain, and $\kappa$ is the anisotropy parameter. In Eq.\ \eqref{pxpham}, $w$ represents the amplitude of spin-flip and $P_j= (1-\sigma_j^z)/2$ denotes projection operators which ensure that two neighboring spins do not simultaneously occupy spin-up states. These projection operators encodes the strong van der Waals interaction between the atoms in their Rydberg state \cite{exp1,ss1,ss2}. For both Hamiltonians, $\lambda(t)$ represents a time-dependent magnetic field. In what follows, we drive the magnetic field periodically by a square-pulse protocol where $\lambda(t)= -\lambda_0$ for $  0\le t \le T/2$ and $\lambda(t)= +\lambda_0$ for $ T/2<t \le T$, with $T= 2\pi/\omega_D$ denoting the time-period. For both the ${\rm XY}$ [Eq.~\eqref{xyham}] and PXP [Eq.~\eqref{pxpham}] spin chains, we have considered periodic boundary conditions.
    
    {\it Periodic dynamics}: The stroboscopic dynamics of such driven chains can be understood in terms of unitary evolution operators $U_a(T,0)\equiv U_a$, where $a=1(2)$ corresponds to the XY (PXP) chain given in Eq.~\eqref{xyham} and Eq.~\eqref{pxpham} respectively. For a square-pulse protocol, one finds
    \begin{eqnarray}
    U_a &=&  e^{-i H_a(\lambda_0) T/(2\hbar)} e^{-i H_a (-\lambda_0) T/(2\hbar)}= e^{-i H^F_a T/\hbar}\,, \nonumber\\
     \label{ueq1}
    \label{eq:hf}
    \end{eqnarray}
    where $H^F_a$ denotes the Floquet Hamiltonian. The state of the driven chain, after $m$ cycles of the drive, is obtained using $|\psi_a(mT)\rangle = U_a(mT,0) |\psi_a(0)\rangle$ where $|\psi_a(0)\rangle$ is the initial state. For the rest of this work, we choose $|\psi_a(0)\rangle= |\cdots \downarrow \downarrow \cdots\rangle$. The stroboscopic dynamics is therefore controlled by the Floquet Hamiltonian $H^F_a$ given in Eq.~\eqref{eq:hf}. 
    
    The evolution operator $U_a(mT,0)$ and hence $H_a^F$, for finite spin chains can be computed numerically by exact diagonalization (ED) of $H_a[\pm\lambda_0]$ yielding $H_a[\pm \lambda_0] |p_a^{\pm}\rangle = \epsilon_{p_a}^{\pm} |p_a^{\pm}\rangle$, where $H_a[+\lambda_0 (-\lambda_0)]$ denote instantaneous Hamiltonians during the positive (negative) part of a drive cycle. In terms of the these eigenvalues and eigenvectors, one finds 
    \begin{eqnarray} 
    U_a &=& \sum_{p_a^+ p_a^-} c_{p_a^+ p_a^-} \exp[-i (\epsilon_{p_a}^+ + \epsilon_{p_a}^-)T/(2\hbar)] |p_a^+\rangle |p_a^-|\, , \nonumber\\ \label{evol1}
    \end{eqnarray}
    where $c_{p_a^+ p_a^-} =\langle p_a^+|p_a^-\rangle$. A numerical diagonalization of $U_a$ yields its eigenvalues $e^{i\theta_{\ell_a}(T) T/\hbar}$ and eigenvectors $|\theta_{\ell_a}(T)\rangle$. This allows us to write $|\psi_a(mT)\rangle = \sum_{\ell_a} \alpha_{\ell_a} e^{im \theta_{\ell_a} T/\hbar} |\theta_{\ell_a}\rangle$, where $\alpha_{\ell_a}=\langle\theta_{\ell_a}|\psi_a(0)\rangle$. 
    
    
    These exact numerical studies can be supplemented by an analytical computation of the Floquet Hamiltonian. In the large drive amplitude regime, this can be carried out using Floquet perturbation theory (FPT) \cite{rev12}. For the XY chain, the analysis  requires fermionization of spins using the Jordan-Wigner transformation: $\sigma_j^{+} = (\sigma_j^-)^{\dagger} = c_j^{\dagger}\ (\prod_{l=1}^{j-1}\exp[ i\pi \sigma_j^z])$ and $\sigma_j^z= 2c_j^{\dagger} c_j-1$, where $c_j$ denotes the fermionic annihilation operator at site $j$. The resulting quadratic Hamiltonian can be written as ${\mathcal H}= \sum_{k>0} \psi^{\dagger} {\mathcal H}_k \psi_k$, where $\psi_k = (c_k,c_{-k}^{\dagger})^T$ are two-component fermionic field operator in momentum space and ${\mathcal H}_k= \tau^z (\lambda(t)-J \cos k) -\tau^x \kappa J \sin k$. Here $\tau^z$ and $\tau^x$ denote the usual Pauli matrices in particle-hole space, and we have set the lattice spacing of the chain to unity. For all calculations of XY chain, we consider the thermodynamic limit $L\to\infty$, i.e.,  $dk \rightarrow0$ while employing the Fourier transform~\cite{supp}.
    
    To obtain a perturbative expression for the Floquet Hamiltonian, we work in the regime $\lambda_0 \gg J$ which allows us to write the zeroth order evolution operator as $U_0(t,0) = \exp[-i \lambda_0 f(t)\sum_{k>0} \tau^z/\hbar]$ where $f(t)= t$ for $0\le t \le T/2$ and $f(t)=T-t$ for $T/2< t \le T$. The first-order term is then computed using $U_1(T,0) =  (-i/\hbar) \int_0^T dt U_0^{\dagger}(t,0) H_1 U_0(t,0)$, where $H_1 = \sum_{k>0} J (-\tau^z \cos k  + \tau^x \kappa \sin k)$. Using $H_{F}^{(1)} = i \hbar U_1(T,0)/T$, this leads to \cite{supp} 
    \begin{eqnarray}
    H_{F}^{(1)} &=& -J \sum_{k>0} (\tau^z \cos k - \frac{\kappa \sin \alpha}{\alpha} (\tau^+ e^{i \alpha} + {\rm h.c.})\sin k)\,, \nonumber\\  \label{xyfl1}   
    \end{eqnarray} 
    where $\alpha= \lambda_0 T/(2\hbar)$. 
    
    A similar calculation can be carried out for the PXP spin chain. Here it is convenient to work directly in terms of spin variables. For $\lambda_0 \gg w$, one can write $U_0(t,0) = \exp[-i \lambda_0 f(t)\sum_{j} \sigma_j^z/\hbar]$. A straightforward calculation then yields \cite{supp,bhas1,bhas2} 
    \begin{eqnarray} 
    H_{F}^{(1)} &=& \frac{w \sin \alpha}{\alpha}  \sum_{j}  (\tilde \sigma_j^+ e^{i \alpha} + {\rm h.c.}). \label{pxpfl1} 
    \end{eqnarray} 
    
    Eqs.~\eqref{xyfl1} and \eqref{pxpfl1} allow us to identify the special frequencies $\omega_n^{\ast}= \lambda_0/(n \hbar)$ ($n \in \mathbb{Z}$), for which $\alpha=n \pi$. For these frequencies, $[H_F^{(1)}, \tau^z]=0$ for the XY chain, while $H_F^{(1)}$ vanishes for the PXP chain. In both case, spin dynamics of the driven chain is controlled by higher-order terms in the Floquet Hamiltonian, leading to qualitatively different properties of the driven chain at these frequencies. We note that for both the models,  $\omega_n^{\ast}$ may shift by a small amount due to renormalization from higher order terms. This effect is stronger for the driven PXP chain \cite{supp,bhas2}.   
    
    \begin{figure}
    \includegraphics[width=0.49 \columnwidth]{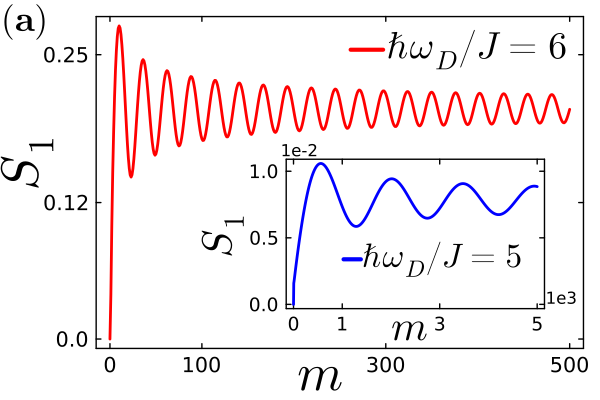}
    \includegraphics[width=0.49 \columnwidth]{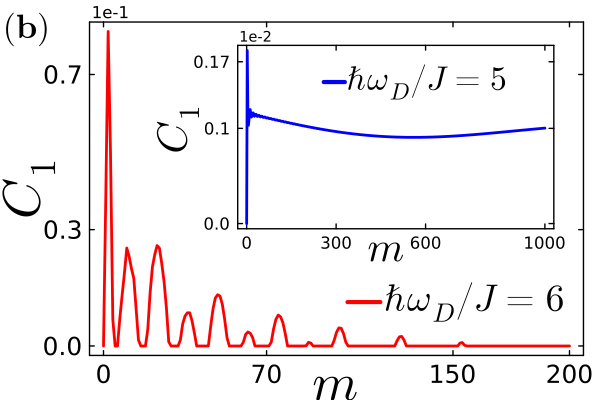}
    \includegraphics[width=0.49 \columnwidth]{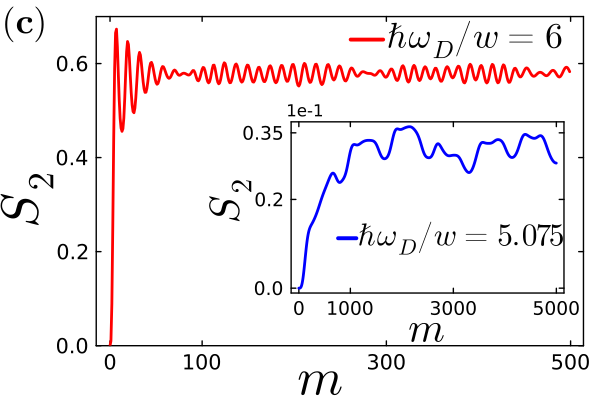}
    \includegraphics[width=0.49 \columnwidth]{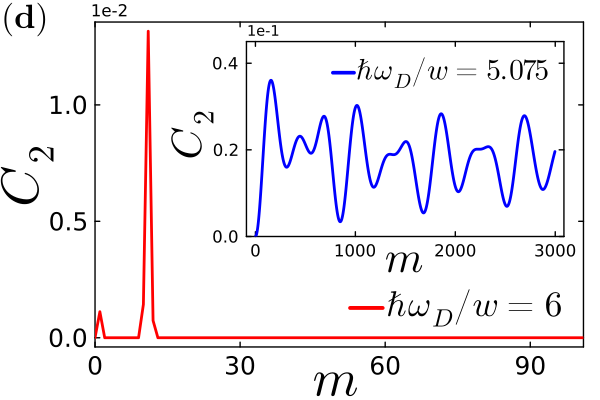}
    \caption{(a) Plot of von-Neumann entropy $S_1(mT)$ as a function of $m$ for the ${\rm XY}$ spin chain driven at non-special frequency  $\hbar \omega_D/J= 6$ without resetting. The inset show plot of $S_1$ vs $m$ for special frequency $\hbar \omega_D/J=5$. (b) Plot of concurrence $C_1(mT)$ as a function of $m$ without resetting for non-special frequency  $\hbar \omega_D/J= 6$. We find that $C_1$ vanishes for large $m$. The inset shows analogous behavior at shifted special frequency $\hbar \omega_D/J=5$ where $C_1$ remains finite for large $m$. We set $\lambda_0/J=10$ and $\kappa=0.7$. (c) and (d) Similar plots but for the driven PXP chain with $\lambda_0/w=10$ and $L=20$ (in $K=0$ and $P=+1$ sector). In all cases discussed in this figure, the system is in a prethermal regime. See text for details.}  \label{fig1}
    \end{figure}
    
    {\it Entanglement and Concurrence}: To compute the entanglement for such a driven state, we first numerically evaluate $\rho_a (mT)= |\psi_a(mT)\rangle\langle \psi_a(mT)|$. Next, following the standard procedure \cite{dsen1,bhas1,bhas2,supp}, we divide the chain into two subsystems $A$ and $B$ of lengths $\ell_A=1$ and $\ell_B=(L-1)$ and compute the partial trace of $\rho_a(mT)$ over the degrees of freedom of subsystem $B$, yielding $\rho_{a}^{\rm red} (mT) = Tr_B \rho_a (mT)$. The von-Neumann entropy $S_a(mT)$ is then obtained from the eigenvalues $\mu_a^i$ of $\rho_a^{\rm red}(mT)$, computed via numerical diagonalization and using $S_a(mT)=-\sum_{i=1}^2 \mu_a^i \ln \mu_a^i$.
    
    A plot of $S_1(mT)$ for the ${\rm XY}$ chain is shown in Fig.\ \ref{fig1}(a). We find that $S_1(mT)$ remains finite and exhibits small oscillations around its steady state value for large $m$ for all drive frequencies. However, for $\omega_D= \omega_n^{\ast}$, $S_1(mT)$ is significantly reduced, and the timescale of these oscillations is much larger, as seen in the inset of Fig.\ \ref{fig1}(a). A qualitatively similar behavior of $S_2(mT)$ is found for the PXP chain [Fig.\ \ref{fig1}(c)].

    In contrast, the concurrence $C_a(mT)$ for such driven chains, obtained from the two-spin density matrices, exhibits a different behavior. To compute $C_a(mT)$, we consider two spins at site $i$ and $j$. The reduced density matrix of these two spins is obtained by tracing out the remaining spins after $m$ drive cycles and can be expressed in terms of equal time correlation functions of the two spins \cite{WW_PRL,Peres_PRL,dsen1}. To this end, we define $\nu_{j}^{1;\alpha} = ((I+\sigma_j^z)/2, (I-\sigma_j^z)/2, \sigma_j^+, \sigma_j^-)$ at site $j$ for the XY chain, where $\sigma_j^{\pm} = (\sigma_j^x\pm i \sigma_j^y)/2$ and $I$ denotes the $2\times 2$ identity matrix. For the PXP chain, one can similarly define $\nu_{j}^{2;\alpha} = ((I+\tilde \sigma_j^z)/2, (I-\tilde \sigma_j^z)/2, \tilde \sigma_j^+, \tilde  \sigma_j^-)$. Using these quantities, the matrix elements of the two-spin reduced density matrix $\rho_{ij}^{(2);a}$ can be written as \cite{WW_PRL,Peres_PRL,dsen1}
    \begin{eqnarray} 
    [\rho_{ij}^{(2);a}(mT)]_{\alpha \beta} &=& \langle \psi_a(mT)| \nu_i^{a;\,\alpha} \otimes \nu_j^{a;\,\beta}|\psi_a(mT) \rangle\,.  \label{twored1}
    \end{eqnarray} 
    The concurrence $C_a(mT)$ can be expressed in terms of the eigenvalues $c_a^i(mT),\,i=1\dots4$ of $\rho_{ij}^{(2);a}(mT) (\sigma_i^y \otimes \sigma_j^y\ [\rho_{ij}^{(2);a}(mT)]^*\ \sigma_i^y \otimes \sigma_j^y)$ as $C_a (mT) = {\rm max} \{0, E_a(mT) \}$, where 
    \begin{eqnarray} 
    E_a(mT)  &=& \sqrt{c_a^1(mT)} - \sum_{j=2}^4\sqrt{c_a^j(mT)}\,.   \label{conc1}
    \end{eqnarray} 
    where we have sorted the eigenvalues $c_a^1 \ge c_a^2 \ge c_a^3 \ge c_a^4$. The matrix elements depends only on $l=|i-j|$. For the XY chain, $C_1(mT)$ decays rapidly with the drive cycle $m$ for $l=1$ and always vanishes in the steady state ($C_1^{st}=0$) due to symmetry reasons~\cite{supp}. For the PXP model, there is a constraint between two consecutive spins. Consequently, it is not meaningful to study the $l=1$ case, since artificial correlations are already induced by this constraint. We thus focus on $l=2$ for both models, since $C_a(mT)$ generally decreases with increasing $l$. 
    
    The behavior of $C_a(mT)$ for the driven XY and PXP chains are shown in Fig.\ \ref{fig1}(b) and (d) respectively. For the XY chain, we find that $C_1 (mT)$ decays rapidly with $m$ for generic drive frequency. In contrast, it retains a small non-zero value for much larger $m$ at $\omega_D=\omega_n^{\ast}$ where higher order terms of the Floquet Hamiltonian control the dynamics [inset of Fig.\ \ref{fig1}(b)]. A similar behavior is observed in Fig.\ \ref{fig1}(d) for the PXP chain. However, we find that $C_2(mT)$ decays very rapidly with $m$ away from the special frequencies and it survives for large $m$ only at the shifted special frequencies. We stress that these features pertain to the prethermal regime. In the infinite temperature steady state of the PXP model, correlations between spins vanish and $C_2^{\rm st} \to 0$ as discussed in the SM~\cite{supp}.

    \begin{figure}
    \rotatebox{0}{\includegraphics*[width=0.492 \columnwidth]{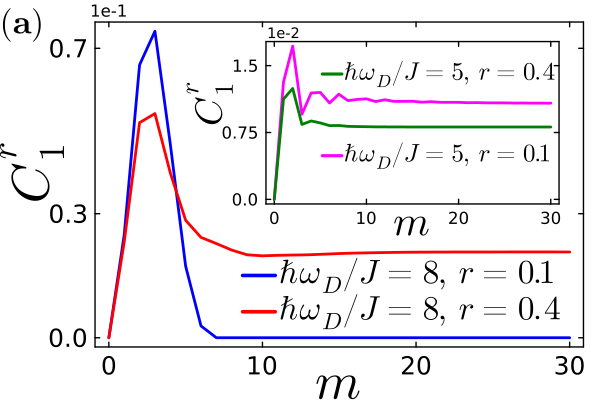}}
    \rotatebox{0}{\includegraphics*[width=0.492 \columnwidth]{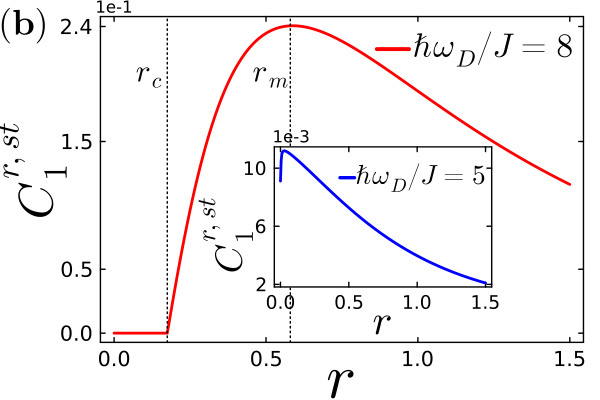}}
    \rotatebox{0}{\includegraphics*[width=0.492 \columnwidth]{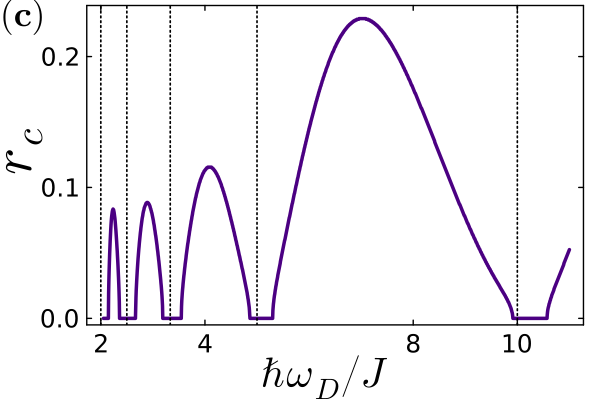}}
    \rotatebox{0}{\includegraphics*[width=0.492 \columnwidth]{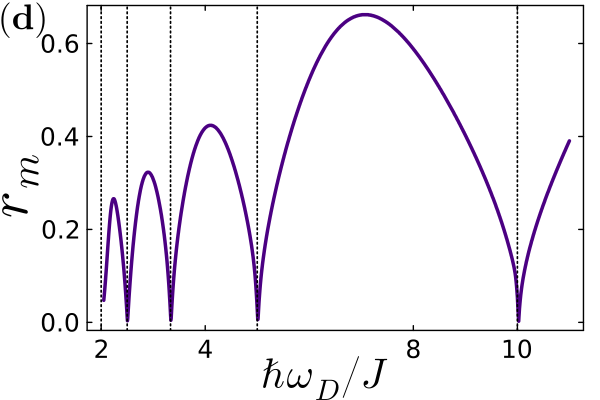}}
    \caption{Plots for the XY model under stochastic resetting. (a) Plot of reset averaged concurrence $C_1^r(mT)$ as a function of $m$ for $\hbar \omega_D/\lambda_0= 0.8$ and $r=0.1$ (blue) and $0.4$ (red) . The inset shows analogous plots at special frequency $\hbar\omega_D/\lambda_0= 1/2$. (b) Plot of steady state concurrence under resetting, $C_1^{r,\rm st}$, as a function of $r$ for $\hbar \omega_D/\lambda_0= 0.8$. The plot shows existence of a critical and optimal resetting rates ($r_c$ and $r_m$ respectively), shown with vertical dotted lines. The inset shows plot of $C_1^{r,\rm st}$ at special frequency $\hbar \omega_D/\lambda_0=1/2$  where $C_{1}^{r,\rm st}$ always remains finite rendering $r_c=0$. (c) Plot of $r_c$ as a function of $\hbar \omega_D/J$. $r_c$ vanishes around $\omega_D=\omega_n^{\ast}$ (shown as dotted vertical lines). (d) Same as in (c) but for the optimal resetting rate $r_m$ which dips around $\omega_D=\omega_n^{\ast}$. We set $\lambda_0/J=10$ and $\kappa=0.7$. See text for more details.  \label{fig2}}
    \end{figure}
    
    {\it Dynamics with stochastic resetting}: We now introduce stochastic resetting in addition to the unitary quantum evolution of our periodically driven system with period $T$. We assume that at the end of each cycle of duration $T$, the system resets back to the initial state $|\psi(0)\rangle$ with probability $0\le p_r\le 1$ and with the complementary probability $(1-p_r)$, the unitary dynamics continues. It is convenient to re-parametrize the reset probability $p_r$ as 
    \begin{equation} 
    p_r= 1- e^{-r T} \, ,
    \label{p_r_def} 
    \end{equation} 
    where the quantity $r$ has the dimension of `rate' or inverse time. Note that $r\in [0,\infty]$. We will loosely refer to $r$ as resetting rate, though the actual resetting also occurs 
    periodically with period $T$. Then the renewal equation for the density matrix reads~\cite{Kusmierz14}
    
    \begin{eqnarray} 
     \rho_a^r(mT) &=&  e^{-r m T}  \rho(mT) +(1-e^{-rT}) \nonumber\\
     && \times \sum_{q=0}^{m-1} e^{-rq T}\ \rho_a(qT)\,.  \label{denres} 
    \end{eqnarray}
    The reset-averaged spin correlators can be then obtained as $\langle \mu_i^{a;\,\alpha} \otimes \mu_j^{a;\,\beta}\rangle = {\rm Tr} [\rho_a^r \mu_i^{a;\,\alpha} \otimes \mu_j^{a;\,\beta}]$. These correlators serve as the matrix elements of the reset-averaged two-spin reduced density matrix. We note that for the steady state where $mT \gg 1$, only the second term of Eq.~\eqref{denres} contribute to spin-correlation function. The reset-averaged concurrence $C_a^r(mT) $ is then obtained from the eigenvalues $c_a^r(mT)$ of this density matrix using Eq.~\eqref{conc1}.

     A plot of $C_1^r(mT)$ for the driven XY chain is shown in Fig.\ \ref{fig2}(a). For generic drive frequencies and $m>10$, $C_1^r(mT) $ vanishes at small $r =0.1$, but saturates to a finite value at larger $r=0.4$, indicating the presence of a finite critical resetting rate $r_c (\sim 0.175)$ above which the steady state concurrence, $C_1^{r,\rm st}$, is non-zero [Fig.\ \ref{fig2}(b)]. In contrast, at the special frequencies $\omega_n^{\ast}$, $C_1^r(mT)$ remains finite at large $m$ even when  $r\to 0$ (inset of Fig.\ \ref{fig2}(a)). This implies $r_c=0$ at these frequencies (inset of Fig.\ \ref{fig2}(b)). Also, for generic drive frequencies, $C_1^{r,\rm st}$ vanishes for both $r=0$ and $r\gg1$ indicating the existence of an optimal resetting rate $r_m$ [Fig.\ \ref{fig2}(b)]. Plots of $r_c$ and $r_m$ as function of $\omega_D$ at fixed $\lambda_0/J$ are shown in Fig.\ \ref{fig2}(c) and \ref{fig2}(d) respectively. Both plots exhibit prominent dips at special frequencies $\omega_D=\omega_n^{\ast}$. 
     
\begin{figure}
    \includegraphics[width=0.49\columnwidth]{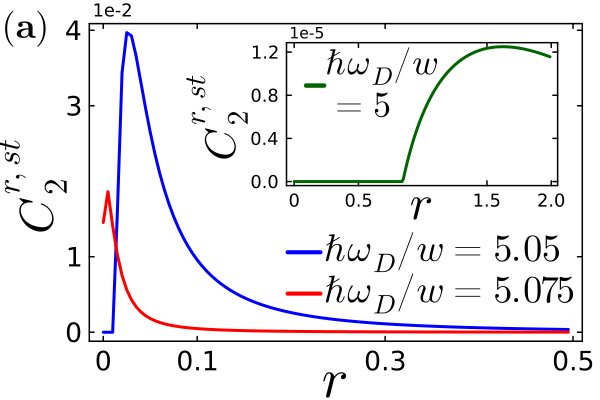}
    \includegraphics[width=0.49\columnwidth]{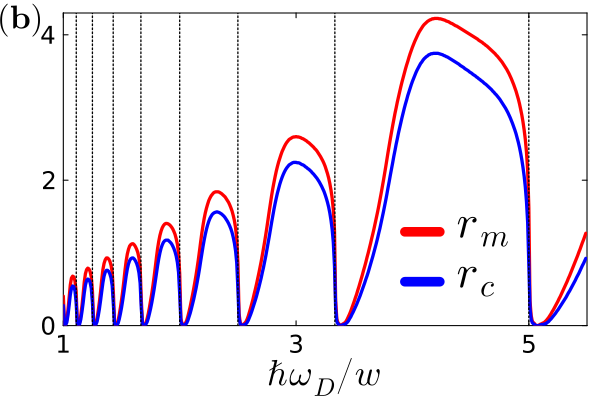} 
    \includegraphics[width=0.49 \columnwidth]{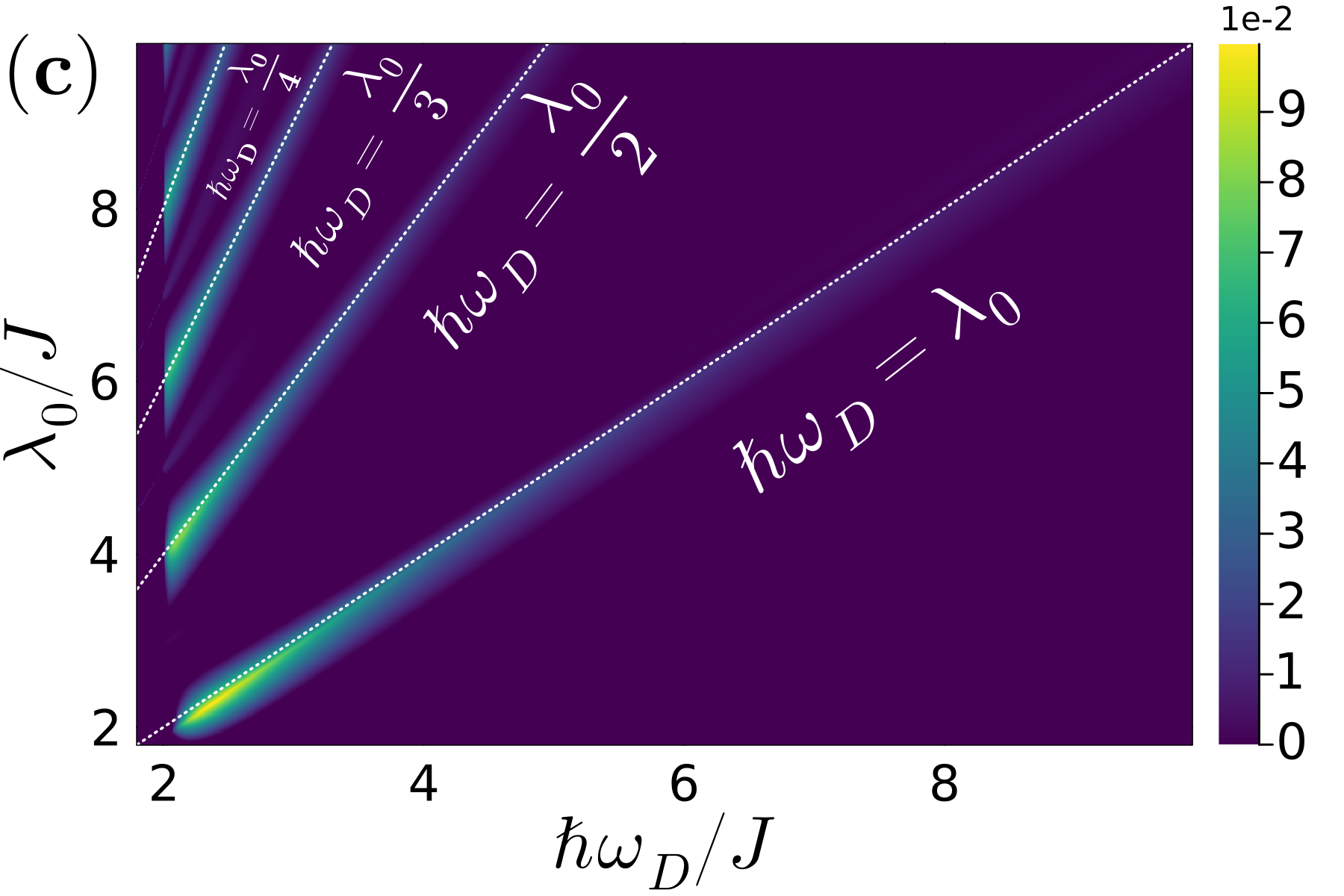}
    \includegraphics[width=0.49 \columnwidth]{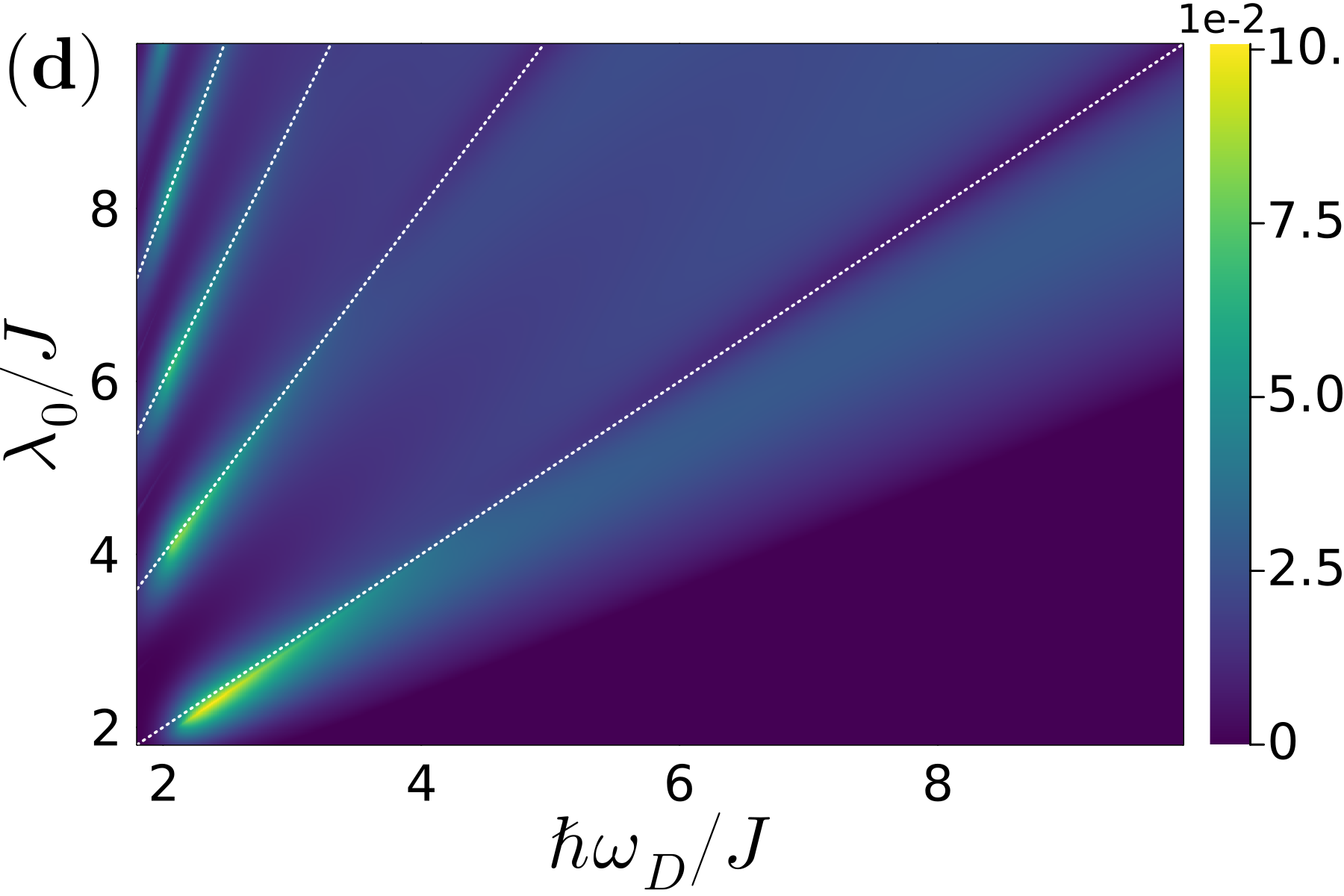}
    \caption{(a) Plot of $C_2^{r,{\rm st}}$ as a function of $r$ for the PXP spin chain driven at the (shifted) special frequency $\hbar \omega_D/\lambda_0= 0.5075$ (red) and at a slightly deviated frequency $\hbar \omega_D/\lambda_0= 0.505$ (blue). Inset shows same for $\hbar \omega_D/\lambda_0= 1/2$ (non-special due to shift). (b) Plot of $r_c$ (blue) and $r_m$ (red) as a function of $\hbar \omega_D/w$, which show dips around slightly right of $\omega_D=\omega_n^{\ast}$ (shown as vertical dotted lines). We set $\lambda_0/w=10$ and $L=20$ (in $k=0$ and $P=+1$ sector). (c) Plot of steady state concurrence $C_1^{\rm st}$ in the absence of resetting, for the XY chain as a function of $\lambda_0/J$ and $\hbar\omega_D/J$ for $\kappa=0.7$. and (d) same plot as (c) for $C_1^{r,{\rm st}}$ at $r=r_m$. See text for details.}
     \label{fig4}
    \end{figure}

    Figures\ \ref{fig4}(a,b) show plots for $C_2^{r,\rm st}$ for the driven PXP chain as a function of $r$. We find that $r_c$ and $r_m$ remain finite over a wide range of frequencies and show qualitatively similar behavior as the XY chain. Both $r_c$ and $r_m$ drops to zero at special drive frequencies. However, in contrast to the XY chain, $C^{r, \rm st}_2$ remains large only near the special frequencies. It drops to much smaller values away from them where $H_F^{(1)}$ governs the dynamics (inset of Fig.\ \ref{fig4}(a)).
    
    The existence of a critical and an optimal reset rates in the concurrence between two spins in both driven integrable and non-integrable systems subjected to stochastic resetting, indicates that these are generic features of such systems. This can be understood as follows. For generic drive frequencies, the concurrence vanishes at large $m$ implying $E_a(mT)<0$ for large $m$. However for finite $r$, the reset averaged spin correlations sample early time dynamics for which $C_a(mT)$ is finite. The weight of these early time contributions increases with $r$, leading to an increase in $E_a(mT)$ at large $m$. Since $C_a^r(mT)= {\rm Max}\{0,E_a(mT)\}$, $C_a^{r,\rm st}$ becomes nonzero only after $r$ exceeds a critical threshold, giving rise to $r_c$. This argument fails when $C_a^{\rm st}\neq 0$ without resetting which occurs within a small window around the special drive frequencies. To explain the existence of an optimal resetting rate $r_m$, we note that, in the limit $r\gg 1$ (the Zeno limit), the system is effectively frozen and remains in the initial state $|\psi_a(0)\rangle$. Since $|\psi_a(0)\rangle$ is chosen to be a Fock state with zero concurrence, $C_a^r(mT)$  and hence $C_a^{r,\rm st}$ vanishes in this limit. Thus, $C_a^{r, \rm st} =0$ for both $r=0$ and $r\gg1$. Given the existence of a finite $r_c$, this mandates that there must be an intermediate reset rate $r_m$ at which $C_a^{r,\rm st}$ attains maxima. 

    {\it Discussion}: Our work is a step forward towards controlled generation and stabilization of entanglement in driven systems, particularly in steady states where heating and decoherence typically suppress quantum correlations. Our analysis of concurrence in both driven integrable and non-integrable spin chains in the presence of a stochastic resetting establishes that the steady state concurrence in the presence of resetting becomes nonzero for $r> r_c$ and attains maxima at $r=r_m$. We also show the existence of special drive frequencies at which $r_c=0$ and $r_m$ attains minima. Such a behavior is tied to the fact that the dynamics at these frequencies is governed by the higher order Floquet Hamiltonians. Our analysis indicates that stochastic resetting can be used as a tool for generating steady state concurrence in driven spin chains. This can be seen explicitly by comparing Figs.\ \ref{fig4}(c) and (d) for the XY chain. The former shows $C_1^{\rm st}=0$ except at $\omega_D=\omega_n^{\ast}$ (lines in the $\lambda_0-\omega_D$ plane) while the latter shows finite $C_1^{r,\rm st}$ over a much wider regime away from the special frequencies. Exploration of this aspect of resetting in other driven quantum systems may be a subject of future study. One way of practically achieving this resetting protocol is as follows. For each sample, one starts from the same initial state and evolves the system unitarily up to an exponentially distributed random time. Thus, each sample has different running time. One can averages an observable over all samples with this `stochastic' running time of the unitary evolution. This is equivalent to sampling configurations from the NESS. Experimentally, one could estimate the full density matrix and from it one can too numerically estimate the two-spin reduced density matrix and then the concurrence from the formula.

    \textit{Acknowledgments}:
    K. S. thanks DST, India for support through the
    project JCB/2021/000030.
    M. K. acknowledges support from the Department of Atomic Energy, Government of India, under project    no. RTI4001. S. N. M. acknowledge support from ANR Grant No. ANR-23-CE30-0020-01 EDIPS. S. G. acknowledges support from CSIR, India under JRF scheme and thanks Tista Banerjee and Saptadip Roy for useful discussions.

    \bibliography{citation}
    \end{document}


\title{Supplementary Material for \enquote{Generating pairwise entanglement in periodically driven quantum spin chains with stochastic resetting}}

\author{Sinchan Ghosh}
\email{sinchanghosh008@gmail.com}
\affiliation{School of Physical Sciences, Indian Association for the Cultivation of Science,
Jadavpur 700032, India}

\author{Manas Kulkarni}
\email{manas.kulkarni@icts.res.in}
\affiliation{International Centre for Theoretical Sciences, Tata Institute of Fundamental Research,
Bengaluru 560089, India}

\author{K. Sengupta}
\email{ksengupta1@gmail.com }
\affiliation{School of Physical Sciences, Indian Association for the Cultivation of Science,
Jadavpur 700032, India}

\author{Satya N. Majumdar}
\email{satyanarayan.majumdar@cnrs.fr }
\affiliation{LPTMS, CNRS, Univ. Paris-Sud, Université Paris-Saclay,
91405 Orsay, France}

\date{\today}
\maketitle

\tableofcontents

\section{Floquet dynamics of driven spin chains}
\label{flsec} 

In this section, we obtain analytical expressions of the Floquet Hamiltonian ($H_F$) for the driven chains. For the square drive pulse protocol that we consider in the main text, $H_F$ can be obtained exactly for integrable XY model. This, along with a perturbative analysis, is discussed in Sec.~\ref{xyfl} . In contrast, for the non-integrable PXP model, we provide only a perturbative analysis of $H_F$ in Sec.~\ref{pxpfl}.

\subsection{XY Model}
\label{xyfl} 

The integrable model Hamiltonian given in Eq.~(1) of main text which we recall here to be
\begin{eqnarray} 
\label{xyham} 
H_1(t) = -\frac{J}{4} \sum_{\langle i j\rangle } ((1+\kappa)\sigma_i^x \sigma_j^x +(1-\kappa) \sigma_i^y \sigma_j^y) 
 -\lambda(t) \sum_j \sigma_j^z,\label{s-xyham} 
\end{eqnarray} 
can be described in term of free fermions using a standard Jordan-Wigner transformation given by
\begin{eqnarray} 
c_j &=& \left(\prod_{\ell <j} \sigma_{\ell}^z \right) \sigma_j^-, \quad \sigma_j^z = (2 c_j^{\dagger} c_j-1) \label{jwt}
\end{eqnarray} 
where $c_j$ denotes the annihilation operator of a fermion in site $j$ of the chain. In terms of these fermions, the Hamiltonian of the XY spin chain can be written as~{\cite{GlenIsing}} (here we simply write $H_1$ as $H$)
\begin{equation}
H(t) = -\frac{J}{2} \sum_{j=1}^{L} \left[ \hat{c}_j^\dagger \hat{c}_{j+1} + \kappa\ \hat{c}_j^\dagger \hat{c}_{j+1}^\dagger + \text{h.c.} \right] - \lambda(t) \sum_{j=1}^{L} \left( \hat{c}_j^\dagger \hat{c}_j - \frac{1}{2} \right) ~ .
\label{q1}
\end{equation}

We note that periodic boundary conditions (PBC) for a spin chain do not directly correspond to those of the associated fermionic chain. The relation between fermionic Hamiltonians with open boundary conditions (OBC) and PBC is given by 
\begin{equation}
H^{PBC}(t) = H^{OBC}(t)+\frac{J}{2} \times e^{i\pi \hat{N}} \times  \left( \hat{c}_L^\dagger \hat{c}_{1} + \kappa\ \hat{c}_L^\dagger \hat{c}_{1}^\dagger + \text{h.c.} \right) ~ .
\label{obc}
\end{equation}
where $\hat{N}=\sum_{j=1}^L \hat{c}_j^\dagger \hat{c}_j$, is the total number operator for fermions. In Eq.~\eqref{obc}, we note that if $\hat{N}$ takes an odd value, we need $\hat{c}_{L+1}=\hat{c}_1$ to ensure that PBC in Eq.~\eqref{obc} is imposed. In other words, Eq.~\eqref{obc} can be defined on a ring. On the other hand,  if $\hat{N}$ is even, we need $\hat{c}_{L+1}=-\hat{c}_1$ to compensate for the extra opposite sign on the $L^{\rm th}$ bond, i.e., last term in Eq.~\eqref{obc}. This leads to anti-periodic boundary conditions (APBC) for fermions. 

For the rest of this work, we focus on 
APBC ensured by $\hat{N}$ taking an even value. We also consider chain length $L$ to be even.
Both $\hat{N}$ taking even values and even $L$ ensure that all $k$-modes remain paired, otherwise unpaired k-modes need to be treated separately. Next, we write the fermionic Hamiltonian in momentum space using 
\begin{equation}
\hat{c}_j = \frac{e^{-i \pi/4}}{\sqrt{L}} \sum_k e^{ikj} \hat{c}_k,\quad \hat{c}_j^{\dagger} = \frac{e^{i \pi/4}}{\sqrt{L}} \sum_k e^{-ikj} \hat{c}_k^{\dagger},
\label{ft}
\end{equation}
where $k$ can take the discrete values $\pm(2n-1)\pi/L$, with $n=1,..,L/2$. Note that Eq.~\eqref{ft} has a global phase compared to a conventional Fourier transform. We then find (we write the hamiltonian with APBC as simply $H$ for the sake of brevity)
\begin{equation}
\begin{aligned}
H(t) = & -\frac{J}{2} \sum_k \left[ 2\cos k\ \hat{c}_k^\dagger \hat{c}_k + \kappa (i\ e^{ik}\ \hat{c}_k^\dagger \hat{c}_{-k}^\dagger + \text{h.c.} )\right] - \lambda(t) \sum_k \left( \hat{c}_k^\dagger \hat{c}_k - \frac{1}{2} \right) ~,\\
=&\sum_{k>0} \begin{pmatrix}
\hat{c}_k^\dagger \quad
\hat{c}_{-k}
\end{pmatrix}
\begin{pmatrix}
(\lambda(t) - J \cos k) & - \kappa J  \sin k \\
-\kappa J \sin k & -(\lambda(t) - J \cos k)
\end{pmatrix}
\begin{pmatrix}
\hat{c}_k \\
\hat{c}_{-k}^\dagger
\end{pmatrix} ~ ,
\end{aligned}
\end{equation}

Using the Pauli matrices $\vec \tau= (\tau^x, \tau^y,\tau^z)$ in the particle-hole space, $H(t)$ can be written as 
\begin{equation}
\begin{aligned}
H(t)= &\sum_{k} \Psi_k^\dagger \, H_k(t) \, \Psi_k, \hspace{5mm} \text{where,}\hspace{3mm} \Psi_k =(\hat{c}_k , \hat{c}_{-k}^\dagger)^T ~ , \\
& \quad \quad H_k(t) = (\lambda(t) - J\cos k)\ \tau^z+ (\kappa J \sin k)\ \tau^x ~.
\end{aligned}
\label{XY}
\end{equation}

We note that the global phase factor used in the Fourier transform [Eq.~\eqref{ft}] of the fermionic fields allows us to write the off-diagonal term in $H(t)$ in terms of $\tau^x$; without them these terms will involve $\tau^y$. The choice of these phase factors does not lead to any other changes and all correlation functions of the model remain identical. The quadratic nature of the Hamiltonian ensures that the wavefunction at any $t$ can be expressed as a tensor product 
\begin{equation}
\ket{\Psi(t)} = \bigotimes_{k>0} \ket{\psi_k(t)},\quad \text{where}\quad \ket{\psi_k(t)} = ( u_k(t) + v_k(t)\ c_{k}^{\dagger} c_{-k}^{\dagger}) \ket{0}  =(u_k(t) \quad v_k(t))^T  ~,
\label{gstate1}
\end{equation}

where $u_k(t)$ and $v_k(t)$ satisfy $|u_k(t)|^2 +|v_k(t)|^2=1$ and are determined by solving the Schrodinger equation 
\begin{eqnarray} 
i\hbar \partial_t \begin{pmatrix}
u_k(t) \\
v_{k}(t)
\end{pmatrix}  =& H_k \begin{pmatrix}
u_k(t) \\
v_k(t)
\end{pmatrix} \,.
\end{eqnarray}

Next, we use $H_k$ to analyze the drive model with a time-dependent magnetic field $\lambda(t)$ given by 
\begin{equation}
    \lambda(t) =
\begin{cases} 
\lambda_0 & \text{for } 0 \leq t \leq \frac{T}{2}, \\
-\lambda_0 & \text{for } \frac{T}{2} < t \leq T.
\end{cases}
\label{p}
\end{equation}

The evolution operator for the driven problem after the end of a drive cycle can be written as $U(T,0)= \prod_{k>0} U_k(T,0)$, where 
\begin{equation}
    \begin{aligned}
    U_k(T, 0) &= e^{-iH_k^{-} T /(2\hbar)}\, e^{-iH_k^{+} T /(2\hbar)} = e^{-iH_{k}^{F}T/\hbar}, \quad
    H_{k}^{\pm} = (\pm \lambda_0 - b_k)\tau^z + \Delta_k \tau^x~.
    \end{aligned}
    \label{fl3}
\end{equation}

The $2\times 2$ matrix structure of $H_k^{\pm}$ allows one to express it in the basis of the Pauli matrices as 
\begin{equation}
\label{eq:nkxyz}
   H_k^F= (\vec{\tau} \cdot \hat{n}_k) |\epsilon_k^F|,
\end{equation}
where $\hat{n}_k$ are unit vectors and $\epsilon_k^F$ denotes the Floquet spectrum (eigenvalues of $H_k^F$). The exact analytical forms of $\epsilon_k^F$ and $\hat{n}_k$ are known~\cite{Sen_EntGen,Banerjee_Asymmetry} and are given by

\begin{eqnarray}
\epsilon_k^F &=& \pm \frac{\hbar}{T}\arccos\left(\cos \phi_k^- \cos \phi_k^+ - (\hat{P}_k^- \cdot \hat{P}_k^+) \sin \phi_k^- \sin \phi_k^+ \right)~, \label{floqSpec} \\
         \phi_k^{\pm} &=&\frac{E_k^{\pm} T}{2\hbar}, \quad E_k^{\pm} = \sqrt{(\pm \lambda_0 - b_k)^2 + \Delta_k^2} ~, \quad
        \hat{P}_k^{\pm}= \left( \frac{\Delta_k}{E_k^{\pm}}, 0 , \frac{\pm \lambda_0 - b_k}{E_k^{\pm}}\right) 
    \label{floquet} \\
    n_{kx} &=& -\frac{2 \lambda_0 \Delta_k}{\sin(T |\epsilon_k^F|/\hbar)} \frac{\sin \phi_k^- \sin \phi_k^+}{E_k^{+}E_k^{-}}, \quad  n_{ky} = \frac{\Delta_k}{\sin(T |\epsilon_k^F|/\hbar)} \sum_{l=\pm} \frac{\sin \phi_k^l \cos \phi_k^{\bar{l}}}{E_k^l}, \label{nks}\\
      n_{kz} &=& \frac{1}{\sin(T |\epsilon_k^F|/\hbar)} \sum_{l=\pm} \frac{(l \lambda_0 - b_k)\sin \phi_k^l \cos \phi_k^{\bar{l}}}{E_k^l} \label{nksz} \,,
\end{eqnarray}
where $n_{kz},n_{kz},n_{kz}$ are components of the unit vector $\hat{n}_k$ that appears in Eq.~\eqref{eq:nkxyz}, $b_k =J \cos k$, $\Delta_k=\kappa J \sin k$, and $\bar{l}=\mp$ for $l=\pm$.

Using Eqs.~\eqref{floqSpec}, \eqref{floquet}, \eqref{nks} and \eqref{nksz}, it is possible to compute the correlators of the driven model. The quadratic nature of the model mandates that there are two independent correlation functions that, after $m$ drive cycles, are given by 
\begin{eqnarray}
    C_d(mT) &=&  \langle \psi_k(mT)| c_k^\dagger c_k|\psi_k(mT) \rangle = |v_k(mT)|^2, \quad 
        C_o(mT) =    \langle \psi_k(mT)| c_k c_{-k}|\psi_k(mT) \rangle =  u_k^*(mT)\ v_k(mT),\nonumber \\ 
    \label{cor1}
\end{eqnarray}
where subscripts `$d$' and `$o$' indicate diagonal and off-diagonal respectively. We also recall that $\psi_k(mT)$ is given in Eq.~\eqref{gstate1}. A straightforward computation using Eq.~\eqref{fl3} and Pauli matrix algebra show that both correlators can be written as 
\begin{equation}
    C_{d,o}(mT) = C_{d,o}^{{\rm GGE}} + \, C_{d,o}^{{\rm odd}}\ \sin(2\, \chi_k(mT)) + \, C_{d,o}^{{\rm even}}\ \cos(2\, \chi_k(mT))~,
    \label{Exact_corr}
\end{equation}
where $\chi_k(mT)= m \epsilon_k^F T/\hbar$. In Eq.~\eqref{Exact_corr}, GGE implies Generalized Gibbs Ensemble which describes the steady state of these integrable models. $C_{d,o}^{\rm GGE}$ gives the corresponding steady state values of the correlators, and $C_{d,o}^{\rm odd (even)}$ indicates parts of the correlators which are odd (even) functions of $\epsilon_k^F$. Using Eqs.~ \eqref{floquet}, \eqref{nks}, and \eqref{cor1}, we obtain 
\begin{eqnarray}
      C_d^{GGE} &=& \frac{1}{2}[(v_k^0)^2 + (n_{ky}u_k^0)^2 + (n_{kz}v_k^0+n_{kx}u_k^0)^2],\quad 
        C_d^{\rm even}=\frac{1}{2}[(v_k^0)^2 - (n_{ky}u_k^0)^2 - (n_{kz}v_k^0+n_{kx}u_k^0)^2]  ~ \nonumber\\  C_d^{\rm odd} &=& -u_k^0\ v_k^0\ n_{ky}, \quad 
    C_o^{GGE} = \frac{1}{2}[((v_k^0)^2 - (u_k^0)^2 )\ n_{kz}\ (n_{kx} - in_{ky}) + v_k^0\ u_k^0\ n_{kx}\ (n_{kx}-2in_{ky})], \nonumber\\
    C_o^{\rm odd} &=& \frac{1}{2}[((v_k^0)^2 - (u_k^0)^2 )\ n_{ky} + i(n_{kx}\ (v_k^0)^2 - 2n_{kz}\ v_k^0\ u_k^0 - n_{kx}\ (u_k^0)^2)] \nonumber\\
     C_o^{\rm even} &=& \frac{1}{2}[-((v_k^0)^2 - (u_k^0)^2 )\ (n_{kx}-in_{ky})+ v_k^0\ u_k^0\ (n_{kz}^2-n_{kx}^2+2i\ n_{kx}\ n_{ky})]~,
    \label{CorrA}
\end{eqnarray}
where $(v_k^0 , u_k^0)$ correspond to the initial state ($t=0$) which is in our case given by $u_k^0=1 , v_k^0=0 \,\, \forall k$. Eq.\ \eqref{CorrA} provides exact expressions of the equal-time correlators of the driven XY chain which is used to compute the different measures of entanglement in the system.

Next, we provide a perturbative expression of the Floquet Hamiltonian in the large drive amplitude regime which will be useful for obtaining the special drive frequencies $\omega_n^{\ast}$ discussed in the main text. To this end, we use Floquet perturbation theory (FPT) and identify the largest term in the driven instantaneous Hamiltonian [Eq.~\eqref{XY}] as $H_k^0= \lambda(t) \, \tau^z$. The remaining terms in the Hamiltonian are denoted by $H_{k}^1=\Delta_k \, \tau^x - b_k \, \tau^z$ and their effects will be estimated using time-dependent perturbation theory. The unperturbed evolution operator is given by
\begin{equation}
U_0(t, 0) =
\begin{cases}
\prod_{k>0} e^{-i \lambda_0 t \tau^z } & \text{for} \hspace{3mm} t \leq \frac{T}{2}, \\
\prod_{k>0} e^{-i \lambda_0 (T - t) \tau^z}  & \text{for} \hspace{3mm} \frac{T}{2} < t \leq T.
\end{cases}
\label{un}
\end{equation}
The higher order evolution operators can be calculated as follows
\begin{equation}
\begin{aligned}
U_m(T,0)& = (-i/\hbar)^m\ \int_0^T dt_1 H_1^I(t_1)\ \int_0^{t_1}dt_2 H_1^I(t_2)....\int_0^{t_{m-1}}dt_m H_1^I(t_m),  \\
\end{aligned}
\label{fl1}
\end{equation}
where  $H_1^I(t)= U_0^{\dagger}(t,0)H_1\ U_0(t,0)$ and $H_1= \sum_{k>0} H_{k}^1$.

The first order evolution operator is given by $U_1(T,0) = (-i/\hbar) \int_0^T dt H_I(t)$. Using the relation 
\begin{equation}
\begin{aligned}
H_1^I(t)& = U_0^{\dagger}(t,0)\ H_1\ U_0(t,0) = \sum_{k>0} e^{2 i \lambda_0 f(t)/\hbar } \Delta_k\ \tau^x\  - b_k\ \tau^z,
\end{aligned}
\label{hamint}
\end{equation}
where $f(t)=t$ for $t\le T/2$ and $f(t)= (T-t)$ for $t >T/2$. Let us evaluate the following integrals~\cite{bhas1,bhas2}
\begin{equation}
\begin{aligned}
I_1(t) =& \sum_{k>0} \int_0^t H_1^I (t_1) dt_1 = \sum_{k>0} \left[-b_k\ t\ \tau^z\   - \frac{ i \hbar\Delta_k}{\lambda_0}\ \sin(\lambda_0\ t / \hbar) \left( e^{i \lambda_0 t/ \hbar}\ \tau^+ + {\rm h.c.} \right) \right] ~,\\
I_2(t,T) =& \sum_{k>0} \int_{T/2}^t  H_1^I (t_1) dt_1 =  \sum_{k>0} \left[-b_k\ (t-T/2)\ \tau^z  - \frac{  i\Delta_k}{\lambda_0}\ \sin(\lambda_0(t-T/2)/\hbar) \left( e^{-i \lambda_0\ (t-3T/2)/\hbar}\ \tau^+ + {\rm h.c.}\right) \right].
\end{aligned}
\label{un1}
\end{equation}
In terms of these integrals one can write $U_1(T,0)= (-i/\hbar) (I_1(T/2) + I_2(T,T))$. The Floquet Hamiltonian can be then read off from $U_1(T,0)$ as 
\begin{equation}
\begin{aligned}
H_{F}^{(1)}=\frac{i \hbar}{T}U_1(T,0) = & \sum_{k>0} \left[- b_k\ \tau^z + \frac{ 2 \hbar \Delta_k}{\lambda_0 T}\ \sin(\lambda_0 T/2 \hbar)\ \left( e^{i \lambda_0 T /2 \hbar}\ \tau^+ +{\rm h.c.} \right) \right].
\end{aligned}
\label{un2}
\end{equation}
This leads to Eq.~5 of main text. We note that
\begin{equation}
    [H_F^{(1)}, \tau^z]=0,\quad \text{for}\quad \lambda_0 T = 2 n \hbar \pi  \quad (n \in Z)\,.
    \label{eq:sf}
\end{equation}

Eq.~\eqref{eq:sf} yields the special frequencies $\omega_n^{\ast}= \lambda_0/(n \hbar)$. This leads to an emergent $U(1)$ symmetry at this order at the special frequencies. 

The computation of the higher-order terms of the Floquet Hamiltonian is cumbersome but can be done using Eq.~\eqref{fl1}. To compute the second-order term in the evolution operator, we similarly define the following integrals 
\begin{equation}
\begin{aligned}
  I_3(t) &= \int_0^t H_1^I (t_1)\ I_1(t_1)\ dt_1 =  \sum_{k>0} \left[ \frac{b_k^2t^2}{2} +\frac{\hbar^2 \Delta_k^2}{\lambda_0^2}[1-\cos(2\lambda_0t/\hbar)]\right] I -i \left[ \frac{\hbar \Delta_k^2}{\lambda_0^2}\left[\hbar \sin(2\lambda_0t/\hbar)-2\lambda_0t\right] \right] \tau^z  \\
   &\hspace{14em}+i\frac{\hbar\ b_k\Delta_k}{\lambda_0^2} [\lambda_0t\ \cos(\lambda_0t/\hbar) - \hbar \sin(\lambda_0t/\hbar)]\ (e^{i\lambda_0 t/\hbar}\ \tau^{+} + {\rm h.c.} ) ~,\\
  I_4(t,T) &= \int_{T/2}^t  H_1^I (t_1)\ I_2(t_1,T)\ dt_1 \\
  &= \sum_{k>0} \left[ \frac{b_k^2}{8}(2t-T)^2 + \frac{2\hbar^2 \Delta_k^2}{\lambda_0^2} [1 - \cos(\lambda_0(2t-T)/\hbar)]\right]I -i\left[ \frac{2 \hbar \Delta_k^2}{\lambda_0^2}\,[\sin(\lambda_0(2t-T)/\hbar) -\lambda_0(2t-T)]\right] \tau^z \\
    & \qquad \qquad + i \left[\frac{\hbar\ b_k \Delta_k}{2\lambda_0^2} [-2\hbar\  \sin(\lambda_0(2t-T)/\hbar)+ \lambda_0(2t-T)\ \cos(\lambda_0(2t-T)/\hbar)] \right] (e^{-i\lambda_0(t-3T/2)/\hbar}\ \tau^{+} +{\rm h.c.}  )~.\\
\end{aligned}
\label{un3}
\end{equation}
Using these, a cumbersome calculation yields $U_2(T,0)= [U_1(T,0)]^2/2$; consequently, we find  $H_F^{(2)}=0$.

Next, we evaluate the third order term in the Floquet Hamiltonian. To this end, we introduce two integrals 
\begin{equation}
\begin{aligned}
  I_5 \left(t \right) = \int_0^t H_1^I \left(t_1 \right)\ I_3 \left(t_1 \right)\ dt_1 , \qquad I_6 \left(t,T \right) = \int_{T/2}^t H_1^I \left(t_1 \right)\ I_4 \left(t_1, T \right)\ dt_1~.
\end{aligned}
\label{un4}
\end{equation}
The third-order evolution operator can be written in terms of these integrals as  
\begin{equation}
\begin{aligned}
  U_3(T,0) &= (-\frac{i}{\hbar})^3 \left(\int_{0}^{T/2} dt_1 I_1 (t_1) + \int_{T/2}^{T} dt_1 I_2 (t_1, T) \right)
 \times  \left(\int_{0}^{t_1} dt_2\ I_1(t_2)\ \theta(T/2 - t_2) + \int_{T/2}^{t_1} dt_2\ I_2(t_2, T)\ \theta(t_2 - T/2)\right) \\
  & \hspace{10em}  \times  \left(\int_{0}^{t_2} dt_3 I_1(t_3) \theta(T/2 - t_3) + \int_{T/2}^{t_2} dt_3\ I_2 (t_3, T)\ \theta(t_3 - T/2)\right) 
\end{aligned}
\label{un5}
\end{equation}
Using this expression of $U_3$, one obtains 
\begin{equation}
\begin{aligned}
  H_{F}^{(3)}&= \frac{i \hbar}{T}[U_3(T,0) - \frac{1}{2} (U_1 U_2 + U_2 U_1)+ \frac{1}{6} U_1^3(T,0)]= \sum_{k>0} 
  (\lambda_{1k} \ \tau^x - \lambda_{2k} \tau^y + \lambda_{3 k} \tau^z) ~~, 
\end{aligned}
\label{un5H}
\end{equation}
where the coefficients $\lambda_{jk}$ for $j=1,2,3$ are given by 
\begin{equation}
\begin{aligned}
\lambda_{3k}&= -\frac{b_k^3\ T^2}{6\hbar^2} + \frac{b_k\ \Delta_k^2}{3 \lambda_0^3\ T}[\lambda_0 T -3 \hbar \sin(\lambda_0 T/\hbar) + 2\lambda_0 T \cos(\lambda_0 T/\hbar)]~,\\
\lambda_{2k}= \frac{\Delta_k}{6\hbar \lambda_0^3\ T}&[(\hbar^2 \Delta_k^2+ 2b_k^2\ [3 \hbar^2+\lambda_0^2 T^2])\sin(\lambda_0T) + 3\hbar \Delta_k^2\ \lambda_0 T  \cos(\lambda_0 T)-2\hbar^2 \Delta_k^2\ \sin(2\lambda_0T) - 6\hbar\ b_k^2\ \lambda_0T]~,\\
\lambda_{1k}= \frac{\Delta_k}{6\hbar \lambda_0^3\ T}&[2  (3\hbar^2 b_k^2 + b_k^2\ \lambda_0^2\ T^2 + \hbar^2 \Delta_k^2)\cos(\lambda_0T) -3\hbar \Delta_k^2\ \lambda_0 T \sin(\lambda_0 T) -2 \hbar^2 \Delta_k^2 \cos(2\lambda_0T) -(6\hbar b_k^2 -b_k^2\ \lambda_0^2\ T^2)]~.
\end{aligned}
\label{un5p}
\end{equation}
We note that $H_F^{(3)}$ does not commute with $\tau^z$ at the special frequencies; the emergent $U(1)$ symmetry is therefore approximate. However, in the prethermal regime, where $H_F^{(1)}$ controls the dynamics, this approximate emergent symmetry shapes the dynamics of the driven chain. We also notice that $H_F^{(3)}$ leads to renormalization of  $H_F^{(1)}$; however, for large $\lambda_0/J$, numerically we observe that this renormalization is negligible.

\subsection{PXP Model}
\label{pxpfl} 
The PXP model described in the main text [Eq.~(2)] which we recall to be
\begin{equation}
H_2(t) = \sum_j ( w \tilde \sigma_j^x -\lambda(t)  \sigma_j^z), \quad \tilde \sigma_j^x = P_{j-1} \sigma_j^x P_{j+1}. \label{s-pxpham}
\end{equation}

where $P_j=(1-\sigma_j^z)/2$ is the projector for spin at the $j$-th site. This model is an effective low-energy description~\cite{EA_RMP,rev13,rev14,exp1,exp2,exp3,exp4,ss1,ss2} of a chain of Rydberg excitations in the regime where $V_0 \gg w, \Delta \gg V_0/2^6$ where $V_0$ being the van der Waals coefficient and $\Delta$ the detuning of the driving laser from the Rydberg state. In this regime,  the effect of van der Walls interaction, $V(i-j)= V_0/|i-j|^6$, can be approximated by disallowing simultaneous Rydberg excitations at neighboring sites, which is incorporated via the projection operators ($P_j$) in the Hamiltonian. In the spin-language used to describe these atoms, an atom in the ground state at any site of the chain corresponds to a spin-down state while a Rydberg excitation corresponds to a spin-up state. In this language, the constraint amounts to forbidding two consecutive spins for being simultaneously in the spin-up state. This model is known to be non-integrable and here we have used Exact Diagonalization (ED) method to solve for the dynamics of the model for a finite chain. 

For the purpose of ED, we note the following points that have been noted earlier in the literature \cite{bhas1, bhas2}.  First, the Hilbert space of the spin model for large $L$ scales as $\phi^L$, where $\phi=(1+\sqrt{5})/2 \sim 1.618$ is the golden ratio. This allows one to access larger $L$ with same computational resource.
Second, we carry out this diagonalization using periodic boundary conditions and for total momentum $K=0$ and $P=1$ sector (where $P$ denotes parity). This further allows us to access larger chains ($L\le 28$). Third, for the square pulse protocol, we define
\begin{equation}
    H_+ =  \sum_j ( w \tilde \sigma_j^x -\lambda_0  \sigma_j^z)~, \quad H_-= \sum_j ( w \tilde \sigma_j^x +\lambda_0 \sigma_j^z)\,,
    \label{eq:hpm}
\end{equation}
and obtain, using ED, eigenstates of $H_{\pm}$: $H_{\pm} |p_{\pm}\rangle = \epsilon_p^{\pm} |p_{\pm}\rangle$. Using these eigenvalues and eigenvector, we can construct the matrix for $U(T,0)$ as
\begin{eqnarray}
  U(T,0) &=& e^{-i H_- T/(2 \hbar)} e^{-i H_+ T/(2\hbar)} = \sum_{p_+ p_-}  e^{-i(\epsilon_p^+ + \epsilon_p^-)T/(2\hbar)} c_{p_+ p_-} |p_-\rangle \langle p_+|,\quad \text{where}\quad c_{p_+p_-}= \langle p_-|p_+\rangle\,. \label{ueq} 
\end{eqnarray}
A numerical diagonalization of $U(T,0)$ yields its eigenvalues $E_{\ell}= \exp[i \theta_{\ell}]$ and corresponding eigenstates $|\ell\rangle$; the eigenvalues of the Floquet Hamiltonian is then obtained~\cite{Sen_EntGen} using $\epsilon_{\ell}^F = (\hbar/T) \arccos [{\rm Re} [E_{\ell}]]$. 

The perturbative computation of the Floquet Hamiltonian has been carried out in details in Refs.\ \cite{bhas1,bhas2} and proceeds as follows. In the large drive amplitude regime, where $\lambda_0 \gg w$, we treat the drive term exactly to obtain 
\begin{equation}
    U_0(T,0) = \exp[i f(t) \lambda_0 \sum_j \sigma_j^z/\hbar],\quad \text{where}\quad f(t)=t\quad \text{for}\quad  0 \le t\le T/2\quad \text{and}\quad f(t)= T-t \quad\text{for}\quad T/2>t>T.
\end{equation}

The contribution of $H_1 = w\sum_j \tilde \sigma_j^x$ is estimated perturbatively. The first-order contribution is given by 
\begin{equation}
    U_1(T,0) = (-i/\hbar) \int_0^T dt\ U_0^{\dagger}(t,0) H_1 U_0(t,0)\,,
\end{equation}
which leads to \cite{bhas1}
\begin{eqnarray} 
H_F^{(1)} &=& \frac{i\hbar}{T} U_1(T,0)=- w \frac{\sin\gamma}{\gamma} \sum_j (e^{-i \gamma} \tilde \sigma_j^+ +{\rm h.c}), \quad \gamma= \lambda_0 T/(2\hbar).
\label{forder} 
\end{eqnarray} 
Eq.~(\ref{forder}) has been used in the main text and leads to special frequencies $\omega_n^{\ast} = \lambda_0/n$ ($n \in Z$) at which $H_F^{(1)}$ vanishes. At these frequencies, the dynamics is controlled by the higher order terms. 

It has been shown in Ref.\ \cite{bhas1} that driven PXP chain hosts an exact symmetry which leads to $\{H_F, {\mathcal C_0}\}=0$ where ${\mathcal C}_0 =\prod_j \sigma_j^z$ is the chirality operator. This precludes presence of second (and all even) order terms in the Floquet Hamiltonian. The expression for the third-order term is given in Ref.\ \cite{bhas2} and yields
\begin{eqnarray} 
H_F^{(3)} &=& \sum_j A_0 (\tilde \sigma_{j-1}^+ \tilde \sigma_{j+1}^+ \tilde \sigma_{j}^- - P_{j-1} \tilde \sigma_j^+ - \tilde \sigma_j^+ - \tilde \sigma_j^+ P_{j+1} +{\rm h.c.} ), \nonumber\\
A_0 &=& \frac{w^3 \hbar e^{-2 i \lambda_0 T/\hbar}}{12 i\lambda_0^3 T} \left( e^{6 i \lambda_0 T/\hbar} + 3e^{i\lambda_0 T/\hbar } (1 + 2i \lambda_0 T/\hbar ) 
+ 2(1-3 e^{2i \lambda_0 T/\hbar} ) \right)
\label{torder} 
\end{eqnarray} 
We note that the last three terms in the expression of $H_F^{(3)}$ in Eq.\ \eqref{torder} leads to renormalization of $H_F^{(1)}$. Consequently, they shift the special frequencies at which $H_F^{(1)}$ becomes minimum. These shifts are, however, small in the regime where $w/\lambda_0 \ll 1$; we have confirmed this numerically using ED. 

\section{Density Matrix and Entanglement Calculation}
\label{dme}

In the main text we have discussed two types of entanglement measures. The first one captures the global aspects of entanglement in the system and is quantified by the von-Neumann entropy for a given pure state \cite{Bennett1996,Meyer2002,nielsen_chuang,VV_RMP,BZ2006,L2016}. We specifically use single-spin reduced density matrix to compute $S$. The other measure, given by the concurrence $C$ (or equivalently negativity), quantifies the entanglement between two spatially separated spins and requires computation of the two-spin reduced density matrix \cite{WW_PRL,HW_PRL,W_inf,Coffman2000,Arul2005}. The procedure for this computation is discussed for the XY model in Sec.\ \ref{dmexy} and the for the PXP model in Sec.\ \ref{dmepxp}. 

\subsection{XY Model}
\label{dmexy} 
\begin{figure}[ht]
    \centering
    \includegraphics[width=0.32\linewidth]{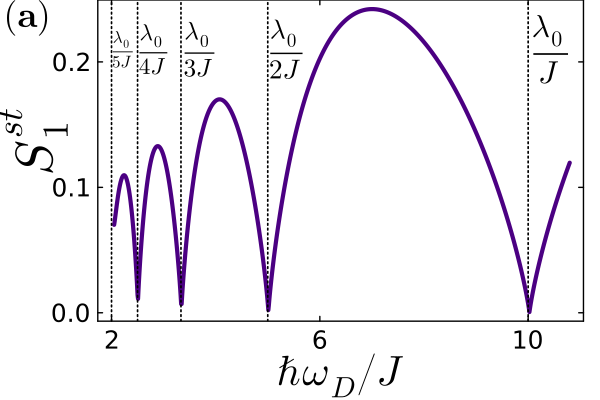}
    \includegraphics[width=0.32\linewidth]{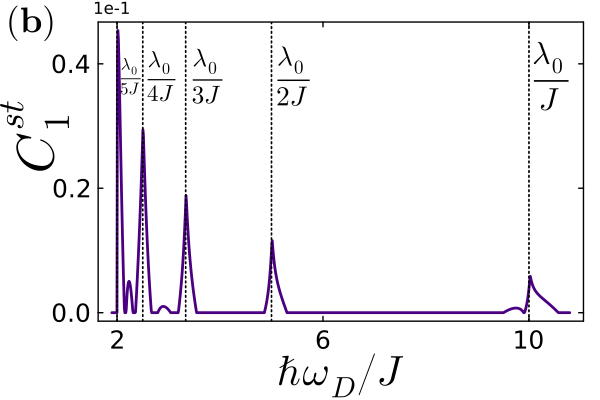}
    \includegraphics[width=0.32\linewidth]{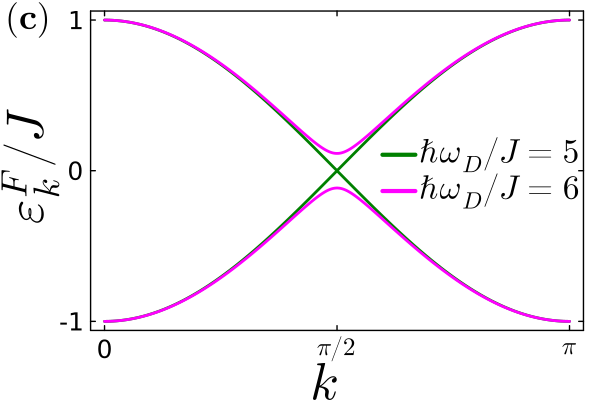}\\
    \includegraphics[width=0.32\linewidth]{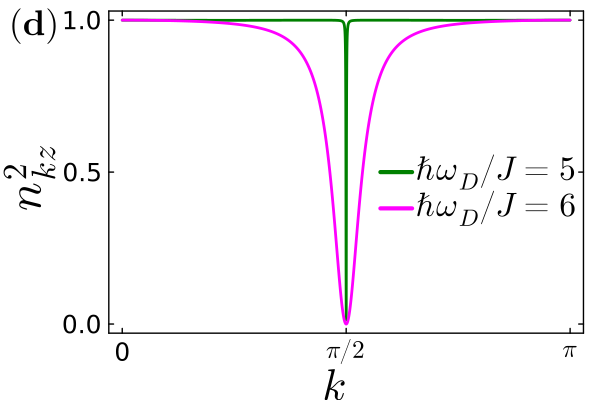}
    \includegraphics[width=0.32\linewidth]{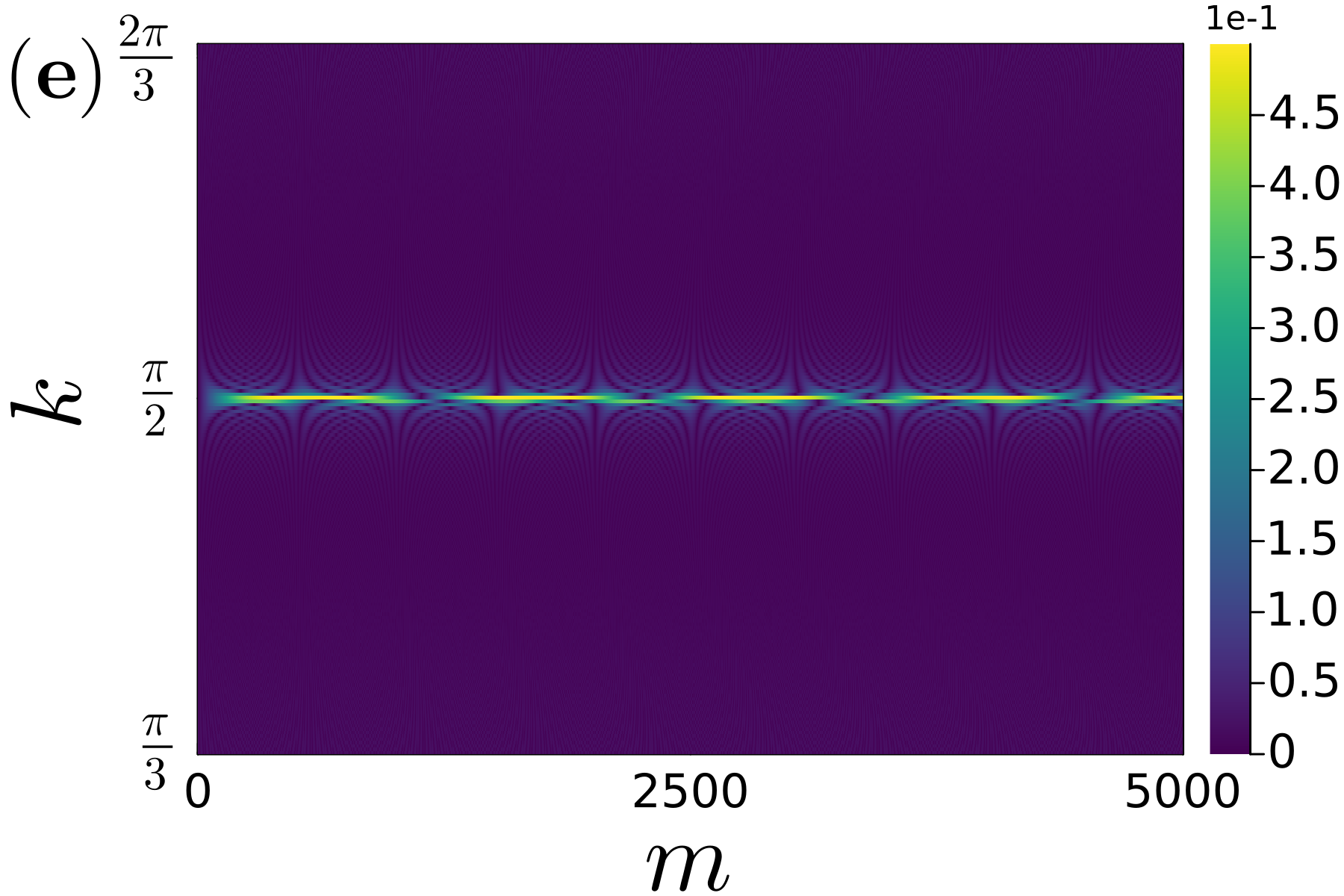}
    \includegraphics[width=0.32\linewidth]{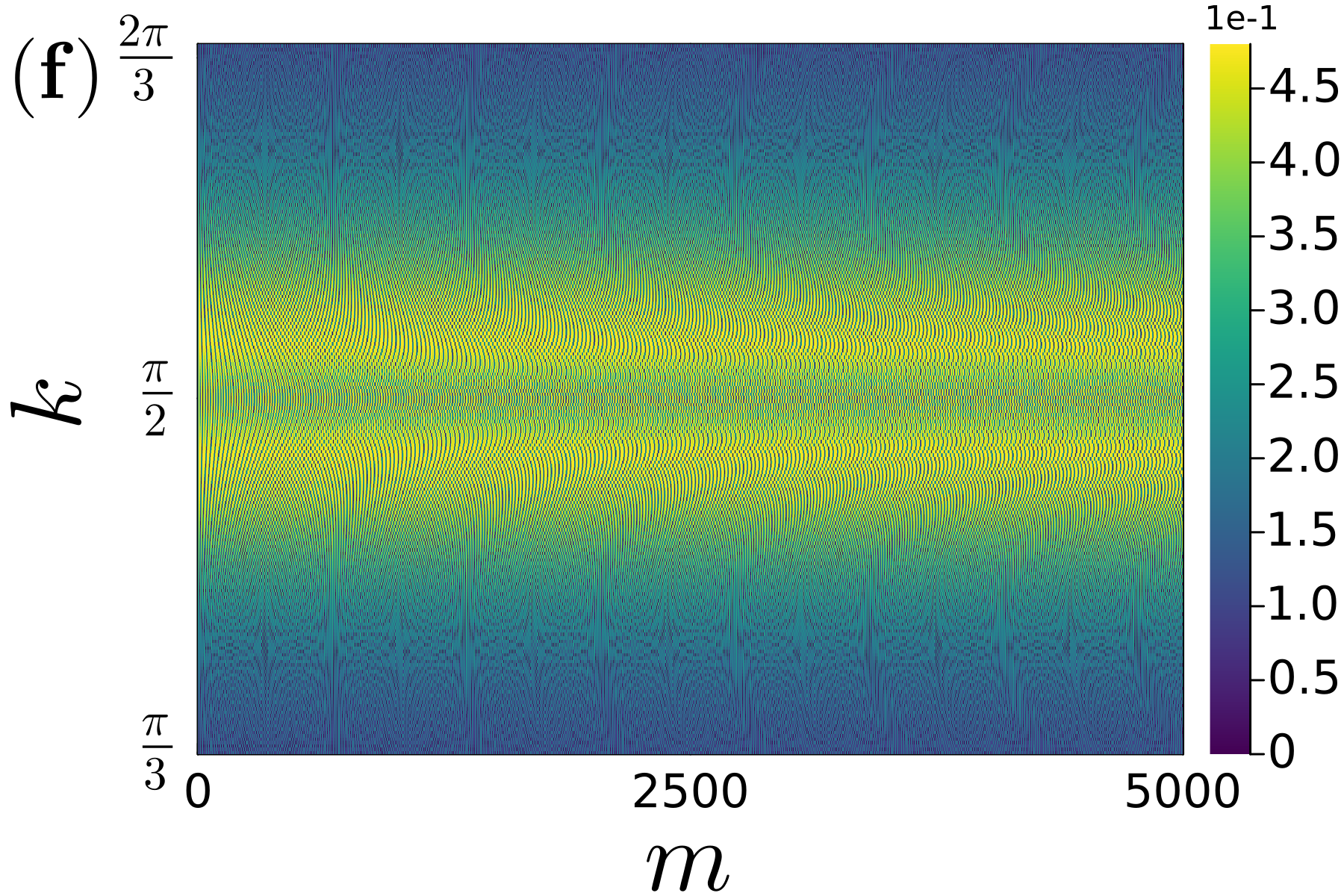}
    \caption{Plots for XY Model. (a) Plot of steady state entanglement ($S_1^{st}$) with drive frequency $\hbar \omega_D/J$. $S_1^{st}$ almost vanishes around $\omega_D=\omega_n^*$. (b) Plot of steady state concurrence ($C_1^{st}$) with drive frequency $\hbar \omega_D/J$. It shows small peaks around $\omega_D=\omega_n^*$. (c) Plot of Floquet quasi-energy bands ($\epsilon_k^F/J$ ) given in Eq.~\eqref{floqSpec} in $1$-st Brillouin zone for a special frequency $\omega_D= \lambda_0/2$ and non-special frequency $\omega_D=3\lambda_0/5$. It shows gap closing for special frequency. (d) Plot of $n_{kz}^2$ for the same pair of frequencies as (c). (e) Heatmap plot of $|C_o|$ in $m-k$ plane for special frequency  $\omega_D= \lambda_0/2$. (f) Heatmap plot of $|C_o|$ in $m-k$ plane for non-special frequency  $\omega_D=3 \lambda_0/5$. We set $\lambda_0/J=10$ and $\kappa=0.7$. See text for more details.}
    \label{XY1}
\end{figure}

 We first calculate the reduced density matrix ($\rho_j$) for a single spin by tracing out other $L-1$ spins of the XY chain. Due to the $\mathcal{Z}_2$ symmetry ($\sigma^x \rightarrow - \sigma^x , \sigma^y \rightarrow - \sigma^y, \sigma^z \rightarrow \sigma^z  $) of the XY model, the off-diagonal elements of $\rho_j$ vanish
 and we obtain 
\begin{equation}
\rho_{j} = 
\begin{pmatrix}
   \frac{1}{2}\langle1 + \sigma^z_j\rangle & \langle \sigma_j^- \rangle \\
   \langle \sigma_j^+ \rangle &  \frac{1}{2} \langle 1 - \sigma^z_j \rangle
\end{pmatrix}
=\begin{pmatrix}
   \alpha_0 & 0 \\
   0 & 1-\alpha_0
\end{pmatrix} ~ .
\label{1-spin}
\end{equation}
 Here we define for any integer $p$    
\begin{eqnarray} 
\alpha_p (mT) &=& \int_{0}^{\pi} \, \frac{dk}{\pi} \, e^{i p k} C_d(mT), \label{alpheq}
\end{eqnarray} 
where we have used, in the thermodynamic limit and with lattice spacing set to unity, $\sum_{k>0} \to (L/\pi) \int_0^{\pi} dk$. Here $C_d(mT) $ is defined in Eqs.\ \eqref{cor1} and \eqref{Exact_corr} and we have used the relation between the $z$ component of the Ising spins with the number operator of the Jordan-Wigner fermions  ($\sigma_j^z= 2\hat n_j-1$) to obtain the matrix elements.  Using Eq.~\eqref{1-spin}, entanglement $S_1(mT)$ is then calculated as
\begin{equation}
\label{supp:s1}
    S_1(mT)= - \alpha_0(mT) \ln [\alpha_0(mT)] - (1-\alpha_0(mT)) \ln [1-\alpha_0(mT)].
\end{equation}
The behavior of $S_1(mT)$ so obtained is discussed in the main text. In Eq.~\eqref{supp:s1}, recall that $m$ denotes the $m$-th cycle at which the observables are computed. The steady state (given by GGE) entanglement, $S_1^{\rm st}$, is  calculated by replacing $C_{d,o}$ with $C_{d,o}^{{\rm GGE}}$ from Eq.~\eqref{CorrA}; this is shown in Fig.\ \ref{XY1}(a) as a function of the drive frequency. In the figure, we find that $S_1^{\rm st}$ shows prominent dips at the special frequencies where the dominant first-order term vanishes and dynamics is controlled by higher order terms in $H_F$.  

Next, we provide details of the concurrence calculation used in the main text. This requires computation of the elements of the two spin reduced density matrix $\rho_{ij}$ (with lattice spacing set to unity). We note that due to $\mathcal{Z}_2$ symmetry of XY model, the off-diagonal elements of $\rho_{ij}$  involving $ \frac{1}{2}\langle(1 \pm \sigma_z) \otimes \sigma_j^{\pm} \rangle $ vanish. This allows one to write \cite{dsen1}
\begin{equation}
\rho_{ij} = 
\begin{pmatrix}
a_{+}^l & 0 & 0 & b_1^{l} \\
0 & a_0^l & b_2^l & 0 \\
0 &  b_2^{l*} & a_0^l & 0 \\
b_1^{l*} & 0 & 0 & a_{-}^l
\end{pmatrix}.
\label{redden}
\end{equation}
where $l=|i-j|$. \par
The diagonal elements of $\rho_{ij}$ can be computed for a general $l$ and are given by 
\begin{equation}
    a_{+}^l  = \alpha_0^2 - \alpha_l^2 + |F_l|^2 ~,\quad
    a_{-}^l  = 1 - 2\alpha_0 + a_{+}^l~,\quad
    a_{0}^l  = \alpha_0 - a_{+}^l ~,
\label{MatDiag}
\end{equation}
Here $ \alpha_l$ in Eq.~\eqref{MatDiag} is defined in Eq.\ \eqref{alpheq} and $F_p$ (for any integer $p$) is defined as 
\begin{eqnarray} 
 F_p (mT) &=& i\int_{0}^{\pi} \, \frac{dk}{\pi} \, e^{ipk}\,  C_o(mT).
 \label{Feq}
 \end{eqnarray}
 where $C_o(mT)$ is defined in Eqs.\ \eqref{cor1} and \eqref{Exact_corr}. 
 
 The computation of the off-diagonal elements are more cumbersome since these correlations involve Jordan-Wigner string operators whose length increases with increasing $l$. In what follows, we therefore restrict ourselves to the case where the two spins are either neighbors ($l=1$) or next-nearest neighbors ($l=2$).  The corresponding off-diagonal elements are given by 
\begin{equation}
\begin{aligned}
b_1^{l=1}= F_1 ~ ,\quad b_2^{l=1}= \alpha_{-1} ~ ,\quad b_1^{l=2}= 2( F_2 \alpha_0 - 2 F_1 \alpha_1) - F_2 ~ ,\quad b_2^{l=2}= 2(\alpha_{-1}^2 -  \alpha_{-2} \alpha_0 + |F_1|^2) + \alpha_{-2}~.
\end{aligned}
\label{MatOffDiag}
\end{equation}
The concurrence can be calculated from $\rho_{ij}$ as shown in main text. It can be readily shown that concurrence will be non-zero only if 
\begin{equation}
    |b_1|>a_0 \quad  \rm or  \quad |b_2|>\sqrt{a_+ a_-}~.
    \label{cond}
\end{equation}
 In what follows, using these conditions, we show entanglement between two consecutive spin vanishes ($C^{l=1}_1=0$) in the steady state of XY model starting from the all spin down initial state ($u_k=1,v_k=0 \,\, \forall k$). Using $C^{{\rm GGE}}_{d,o}$ from Eq.\ \eqref{CorrA} , it can be shown from Eq.\ \eqref{CorrA} that $|v_k|^2$ is symmetric and $u_k^* v_k$ is anti-symmetric under $k \rightarrow \pi -k$ operation for this initial state. From these observations it can be readily shown that for odd integer $l$, $\alpha_l =F_l =0$. This in turn leads to $b_1^{l=1}=b_2^{l=1}=0$ (note that $\alpha_p=\alpha_{-p}$ for $p>0$). Hence the conditions in Eq.~\eqref{cond} never satisfy leading to vanishing concurrence in the steady state.

Next, we consider the concurrence between two spins separated by one site ($l=2$). The corresponding $C_1(mT)$ has been shown in the main text.  A plot of $C_1^{\rm st}$ is shown in Fig.\ \ref{XY1}(b). We find that the steady state concurrence vanishes for a generic drive frequency; in contrast, they remain finite around the special drive frequencies ($\omega_D=\omega_n^{\ast}$). This leads to spikes of finite width in $C_{1}^{\rm st}$ as shown in Fig.\ \ref{XY1}(b).

The dips of $S_1^{\rm st}$ and the spikes of $C_1^{\rm st}$ at the special drive frequencies can be understood as follows.  At these frequencies, the dynamics receives contribution from the higher order terms in $H_F$ and is therefore slower compared to that at generic drive frequencies. Consequently, some entanglement in local parts survives. To understand this, we analyze the Floquet Hamiltonian and find that at $k=\pi/2$, there are exact freezing mode at some specific drive frequencies. At $k=\pi/2$, we get $b_k=0, \Delta_k=\kappa J$ , which leads to $\phi^+=\phi^-=\phi_0$ in Eq.~\eqref{floquet}. Using Eq.~\eqref{nksz}, we find 
\begin{equation}
    n_{kz}|_{k=\frac{\pi}{2}} = \frac{\lambda_0 \sin \phi_0 \, \cos \phi_0}{\Omega~\sin (T\epsilon_{\pi/2}^F/\hbar)  }  \sum_{l=\pm} l  =0, 
\end{equation}
where $\Omega=\sqrt{\lambda_0^2 + \kappa^2 J^2}$. The Floquet evolution operator at $k=\pi/2$ [Eq.~\eqref{fl3}] is therefore given by
 \begin{equation}
     U(T,0)\Big|_{k=\frac{\pi}{2}} =
\begin{pmatrix}
1 + \frac{\kappa^2 J^2}{\Omega^2} (\cos(\Omega T/\hbar)
-1)&
\frac{\kappa J}{\Omega^2} \{ \lambda_0 \big(1 - \cos(\Omega T/\hbar)\big) - i \Omega \sin(\Omega T/\hbar) \}\\
\frac{\kappa J}{\Omega^2} \{-\lambda_0 \big(1 - \cos( \Omega T/\hbar)\big) - i \Omega \sin( \Omega T/\hbar) \}
&
1 + \frac{\kappa^2 J^2}{\Omega^2} (\cos(\Omega T/\hbar)
-1)
\end{pmatrix}
\label{pi/2}
 \end{equation}
From Eq.\ \eqref{pi/2}, we find that for $k=\pi/2$, $U(T,0)=I$ for $\Omega T = 2n \hbar\pi$. This leads to some new special frequencies $\omega'_n= \Omega/(n \hbar)$. Further for $\lambda_0 \gg J$, these frequencies coincide with the special frequencies discussed in the main text: $\omega'_n \simeq \lambda_0/n\hbar = \omega_n^*$. We find the deviation of $\omega'_n$ from $\omega_n^{\ast}$ remains small in the regime considered in this work. In Fig.~\ref{XY1}(c), the Floquet bands show an band gap closing at $k=\pi/2$ for the special frequencies; in contrast, the band gap is prominent at the generic frequencies. Now focusing on the dynamics of the correlators $C_d$ and $C_o$, we get, for our specific initial state,
\begin{equation}
    C_d(mT)= \frac{1}{2}(1-n_{kz}^2)\,\{1- \cos (2m|\epsilon_k|T)\}, \qquad |C_o(mT)|=\frac{1}{2}\sqrt{1-n_{kz}^2} \,\, | -n_{kz}+ e^{-2im|\epsilon_k|T} |
\end{equation}
 We note that $n_{kz}^2$ deviates from unity only at $k=\pi/2$ for special frequencies [Fig.~\ref{XY1}(d)]. This fact along with the presence of freezing mode enforces the correlators to be strictly confined around $k=\pi/2$ at the special frequencies [Fig.\ref{XY1}(e)], whereas the contribution to these correlators are relatively widespread in $k$ space for generic frequencies [Fig.~\ref{XY1}(f)]. This implies that these correlators are delocalized in real space; this leads to stronger bipartite entanglement leading to larger concurrence in the steady state for the special drive frequencies.
 
\subsection{PXP Model}
\label{dmepxp} 

For the non-integrable PXP model, both the single-spin ($\rho_j$) and the two-spin ($\rho_{ij}$) reduced density matrices need to be computed numerically by tracing out the remaining spins. Here we recall from the main text that our driven systems have a large prethermal regime as we remain in perturbative regime. So we cannot reach the thermal state with the dynamics unlike the XY model, where steady state (given by GGE) values were obtained analytically (replacing $C_{d,o}$ by $C_{d,o}^{GGE}$). However we note that for steady state of PXP, which is given by an infinite temperature thermal ensemble, correlations between spins vanish leading to vanishing concurrence ($C_2^{\rm st}=0$). \par

\begin{figure}[ht]
    \centering
    \includegraphics[width=0.32\linewidth]{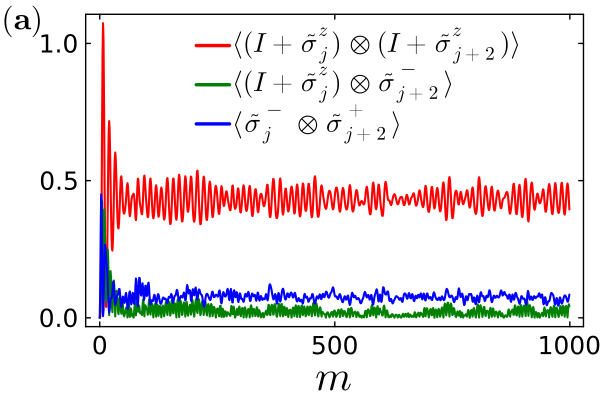}
    \includegraphics[width=0.32\linewidth]{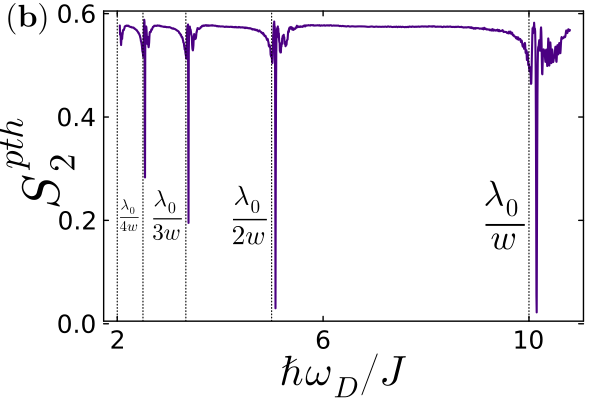}
    \includegraphics[width=0.32\linewidth]{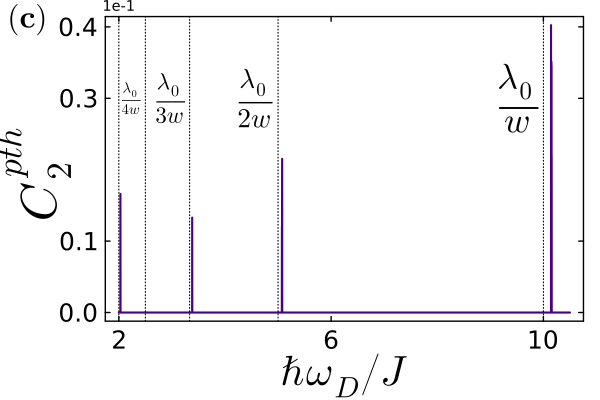}
    \caption{Plots for PXP model. (a) Plot of representative correlators as a function of drive cycle ($m$).The system size has been taken a bit large ($L=22$) to reduce fluctuation. (b) Plot of steady state entanglement ($S_2^{\rm pth}$) with drive frequency $\hbar \omega_D/J$. $S_2^{\rm pth}$ has dips around shifted special frequencies. (c) Plot of steady state concurrence ($C_2^{\rm pth}$) with drive frequency $\hbar \omega_D/J$. $C_2^{\rm pth}$ has very sharp spikes around shifted special frequencies. See text for more details. We set $\lambda_0/w=10$. See text for more details.}
    \label{PXP}
\end{figure}

Unlike the XY model, all the elements of $\rho_j$ and $\rho_{ij}$ can be non-zero in principle since the PXP model does not have similar symmetries as the XY model. However, numerically we find that some of the off-diagonal correlators almost vanish in the prethermal state to give similar prethermal averaged density matrix structure as that for the XY model. The single-spin and the two-spin reduced density matrices for the PXP model, in the prethermal regime, are approximately given by  
     \begin{equation}
         \rho_j^{{\rm pth}}\approx\begin{pmatrix}
        a^{\rm pth} & 0 \\
        0 & b^{\rm pth}
       \end{pmatrix} \quad , \quad 
   \rho_{ij}^{\rm pth} \approx
    \begin{pmatrix}
    a_{+}^{l,{\rm pth}} & 0 & 0 & b_1^{l,{\rm pth}} \\
    0 & a_0^{l,{\rm pth}} & b_2^{l,{\rm pth}} & 0 \\
    0 &  (b_2^{l,{\rm pth}})^* & a_0^{l,{\rm pth}} & 0 \\
    (b_1^{l,{\rm pth}})^* & 0 & 0 & a_{-}^{l,{\rm pth}}
    \end{pmatrix},  \label{pxpred}
     \end{equation}
     where $l=|i-j|$, the lattice spacing is set to to unity and the superscript ``pth" implies prethermal averaged value. To determine these coefficients, we study the behavior of the correlators using exact dynamics; this has been shown in Fig.~\ref{PXP}(a). We find that $\langle (I\pm \tilde \sigma_j^{z}) \otimes \tilde \sigma_{j+2}^\pm \rangle \approx 0$ in the prethermal state, whereas $\langle (I \pm \tilde \sigma_j^{z}) \otimes (I \pm \tilde \sigma_{j+2}^{z}) \rangle$ and $\langle \tilde \sigma_j^\pm \otimes \tilde \sigma_{j+2}^\pm \rangle$ have finite prethermal values and comprise the diagonal and non-vanishing off-diagonal elements respectively. This leads to the structure of the prethermal reduced density matrices in Eq.\ \eqref{pxpred}.

      The prethermal averaged von-Neumann entropy ($S_2^{\rm pth}$) behavior (obtained by $S_2^{\rm pth}=\sum_{m=1001}^{1100} S_2(mT)/100$) for the PXP model with the drive frequency is shown in Fig.~\ref{PXP}(b). It qualitatively matches with the XY model with two significant differences. First, the special frequencies are shifted due to renormalization from higher order; this effect is significantly stronger for the PXP model. Second, the window around the special frequencies where one gets a reasonable concurrence is much smaller. Similarly the prethermal averaged concurrence ($C_2^{\rm pth}$) behavior (obtained by $C_2^{\rm pth}=\sum_{m=1001}^{1100} C_2(mT)/100$) with the drive frequency is shown in Fig.~\ref{PXP}(c). This also shows a qualitative matches with its counterpart for the XY model but with the differences mentioned above.

\section{Reset Averaged Dynamics}
\label{resav}

In this section, we discuss the details of stochastic resetting and the resulting dynamics after resetting. The renewal equation for continuous time resetting protocol has been discussed for quantum systems earler~\cite{Mukherjee_2018,KM_PRA_23}. But, as mentioned in the main text, we need discrete time resetting protocol~\cite{Kusmierz14} for our periodically driven systems because we only observe our systems in discrete stroboscopic times, $mT$ where $T$ is the drive period and $m$ is the number of cycles. We assume that at the end of each cycle of duration $T$, the system resets back to the initial state $|\psi(0)\rangle$ with probbability
$0\le p_r\le 1$ and with the complementary probability $(1-p_r)$ the unitary dynamics continues. However, it is convenient to re-parametrize the reset probability $p_r$ as 
\begin{equation} 
p_r= 1- e^{-r T} \, ,
\label{p_r_def} 
\end{equation} 
where the quantity $r$ has the dimension of `rate' or inverse time. Note that $r\in [0,\infty]$. 
As mentioned in the main text, we will loosely refer to $r$ as resetting rate, though the actual resetting also occurs 
periodically with period $T$.

\begin{figure}[ht]
\centering

    \centering
    \includegraphics[width=0.5\linewidth]{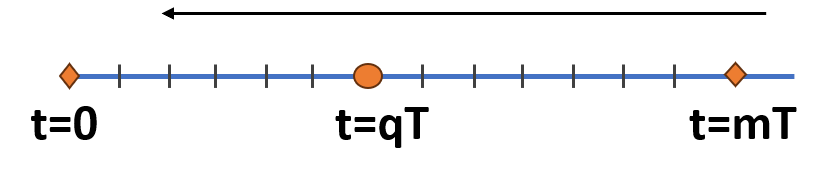}
\hfill
    \caption{Schematic of setup for obtaining reset averaging form. Our observation time is $mT$. Arrow in backward direction implies going backward in time. The point $qT$ (until when no reset happens) can be anywhere between $0$ and $mT$, over which we take the average.}
    \label{reset_sc}
\end{figure}

To obtain the renewal equation, we consider observational drive cycle to be the $m$-th cycle and we consider time in the backward direction. Let us say that no reset happens upto the $q$-th cycle counting backwards from the $m$-th cycle. This can be visualized in Fig.~\ref{reset_sc}.

The probability of no reset upto $q$-th cycle is $(1-p_r)^q $ and the probability of a reset event at q-th cycle is $p_r$. So the conditional probability of a reset event happening at q-th cycle without any reset before (going backward from m-th cycle) is given by \par

\begin{equation}
\begin{aligned}
   p(qT|mT) = (1-p_r)^q\ \delta(m-q) + p_r\ (1-p_r)^q
   =  e^{-rqT}\delta(m-q) +(1-e^{-rT})\ e^{-rqT} \,,
\end{aligned}
\label{eq:s-pqm}
\end{equation}
where to obtain the last part in Eq.~\eqref{eq:s-pqm}, we used Eq.~\eqref{p_r_def}.
This probability distribution is normalized, i.e., $\sum_{q=0}^{m-1}p(qT|mT) = 1$.
After taking stochastic average of the density matrix (varying $q$ for all possible drive cycles) we find the renewal equation for it to be
\begin{equation}
\begin{aligned}
   \rho^r(mT) =  e^{-rmT}  \rho(mT) +(1-e^{-rT})\ \sum_{q=0}^{m-1} e^{-rqT}\ \rho(qT) ~.
\end{aligned}
\label{Renewal}
\end{equation}

\subsection{XY Model}
\label{resavxy} 

For the XY model, instead of taking the average over the reduced density matrix elements, we have taken average over the correlators $C_{d,o}$ which are the building blocks of the matrix elements. The reset averaged forms can be derived using Eq.~\eqref{Exact_corr} and Eq.~\eqref{Renewal} and given by

   \begin{equation}
    \begin{aligned}
    C_{d,o}^r(mT) =C_{d,o}(mT)\, e^{-rmT}\, + \, & (1-e^{-rmT}) \times C_{d,o}^{\text{GGE}}   \,+\, (1-e^{-rT})\times \left\{ C_{d,o}^{\text{odd}}\times  \text{Im}\left[ \frac{1-\exp(-rmT+2imT|\epsilon_k^F|)}{Z}\right] \right.\\
    & \left. \hspace{14em} +\, C_{d,o}^{\text{even}}\times \text{Re} \left[\frac{1-\exp(-rmT+2imT|\epsilon_k^F|)}{Z}\right] \right\} ~,
    \end{aligned}
    \label{ResDyn}
    \end{equation}
   where $Z=1-\exp(-rT+2iT|\epsilon_k|)$ and  the last two terms under curly bracket are obtained using
    \begin{equation}
    \begin{aligned}
        \sum_{q=0}^{m-1} C^{\text{odd}}\, \sin (2iq|\epsilon_k^F|T) \, \exp(-rqT) &=C^{\text{odd}}\,\text{Im}\left[ \sum_{q=0}^{m-1}  \exp (2iq|\epsilon_k^F|T -rqT)\right] \\&=C^{\text{odd}}\,\text{Im}\left[  \frac{\{1-\exp(-rmT+2imT|\epsilon_k^F|)\}}{Z} \right]\, ,
    \end{aligned}
    \end{equation}
     and similarly for the term associated with $C^{\text{even}}$. An interesting thing to notice is that under stochastic reset, the dynamics very quickly (within 20-30 cycles) reaches a steady state (Fig.(3a) of main text). Now, to obtain the NESS values of these correlators we set $m \rightarrow \infty$, and perform the summation to obtain
\begin{equation}
\begin{aligned}
    C_{d,o}^{r , SS} = C_{d,o}^{GGE} \,+\,  (1-e^{-rT})\times \left\{C_{d,o}^{\text{odd}}\times \text{Im}[1/Z] 
     \,+ \, C_{d,o}^{\text{even}}\times  \text{Re}[1/Z] \right\}~.
\end{aligned}
\label{ResSS}
\end{equation}
We have numerically checked that taking average over the density matrix elements leads to qualitatively similar results for large system size.
\subsection{PXP Model}
\label{resavpxp} 
For the reset average calculations, we have taken reset average, using Eq.~\eqref{Renewal}, of the elements of the two spin reduced density matrix, obtained by tracing out the remaining spins numerically.  Similar to the XY model, in this case also, reset averaged dynamics reaches steady state very quickly (within 10-20 cycles).  Using this fact, we have done NESS calculation by staying on a relatively large drive cycle. 

  \begin{figure}[ht]
    \centering
    \includegraphics[width=0.32\linewidth]{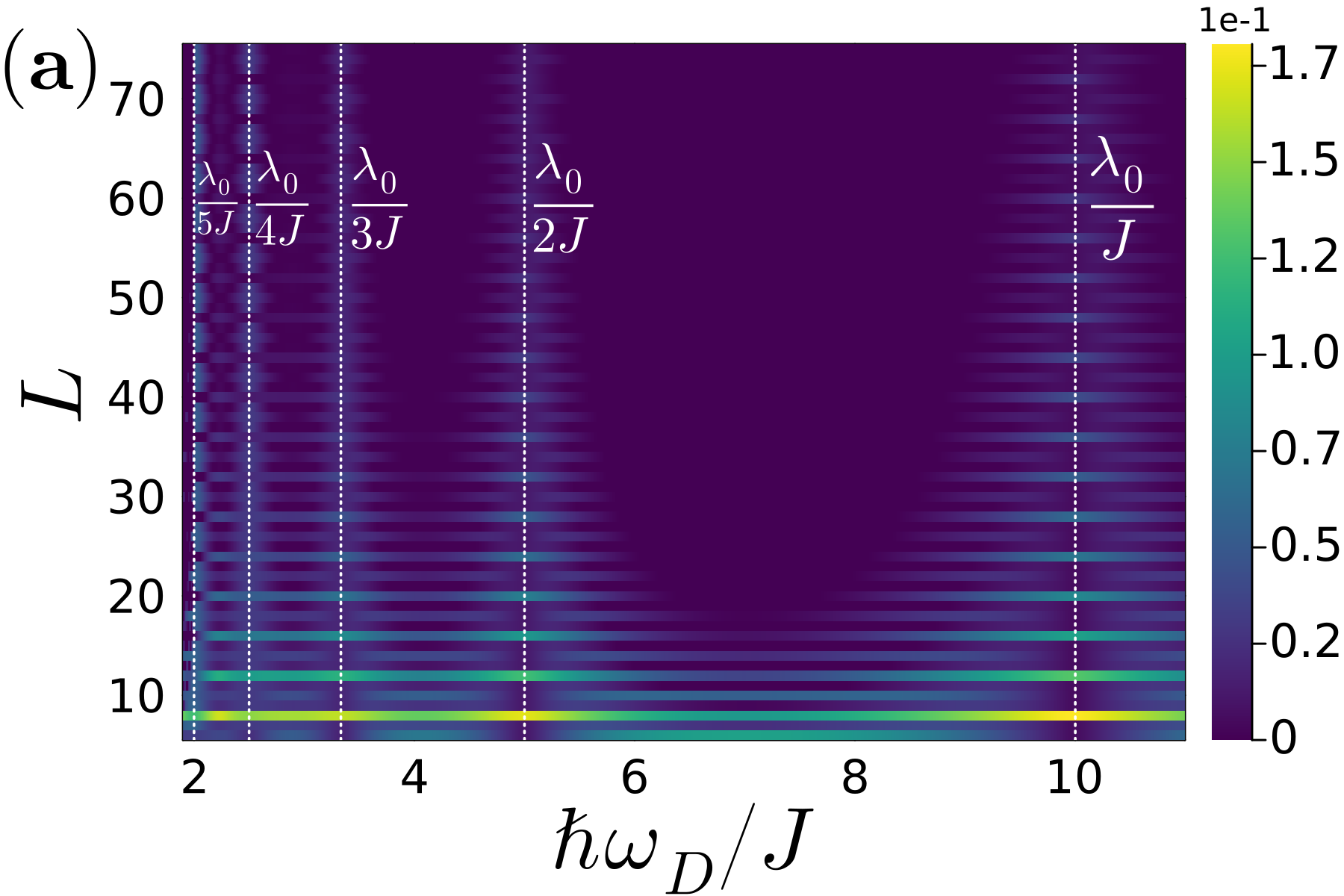}
    \includegraphics[width=0.32\linewidth]{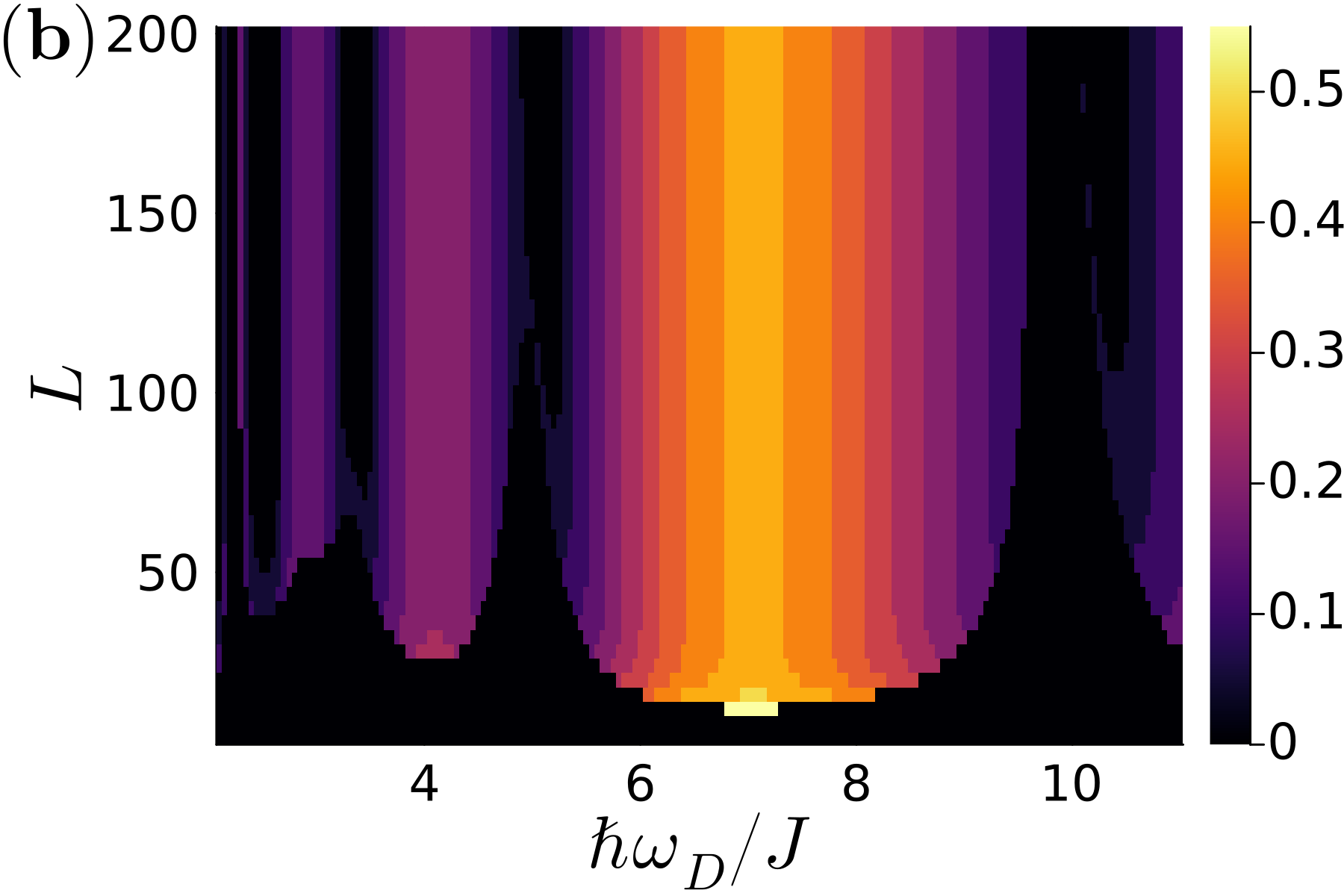}
    \includegraphics[width=0.32\linewidth]{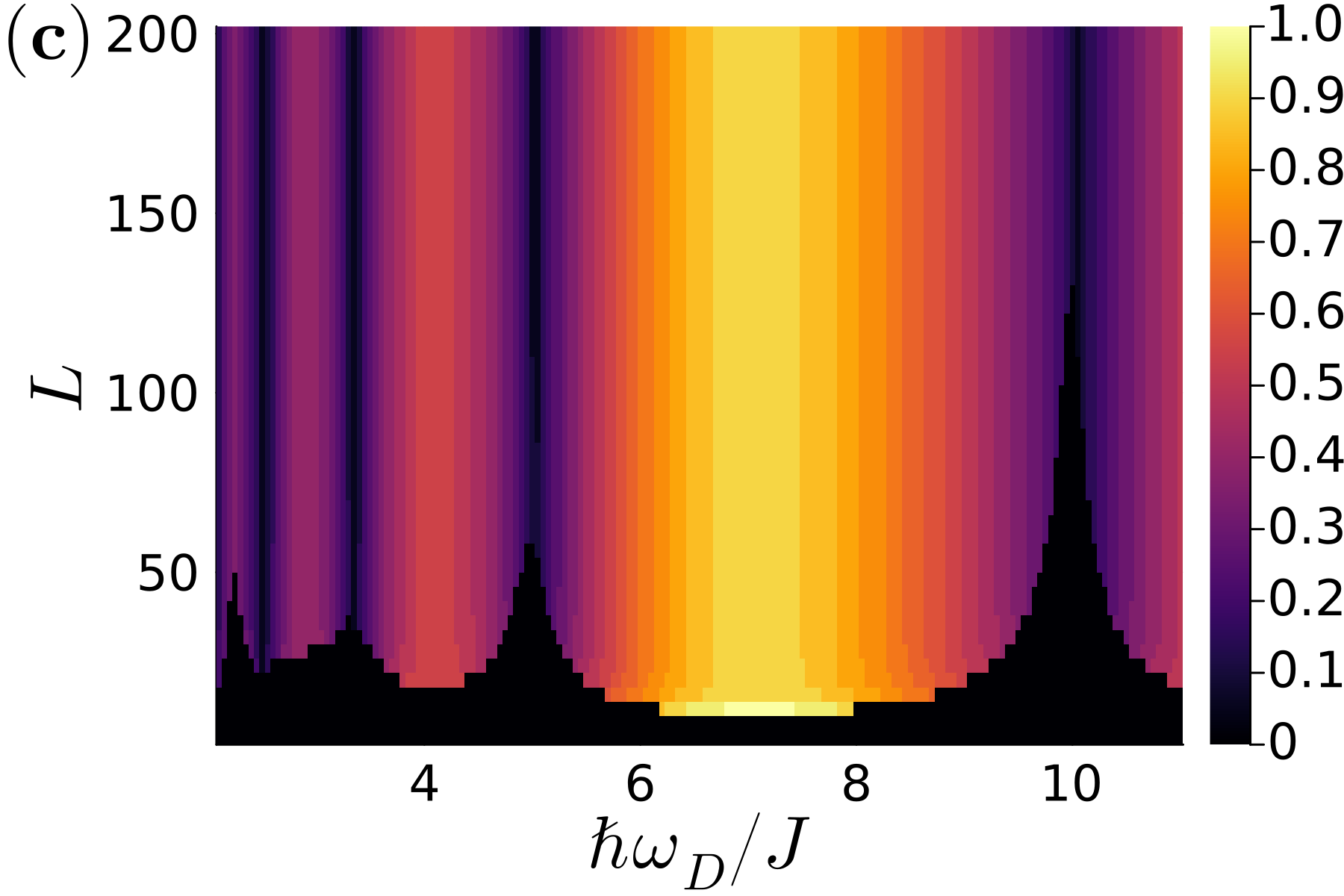}
    \caption{ Heatmap plot for XY model of (a) $C^{st}$  (b) $r_c$ and (c) $r_m$  in $L-\hbar \omega_D/J$ plane. We set $\lambda_0/J=10$ and $\kappa=0.7$.  See text for more details.}
    \label{XY2}
\end{figure}

\section{Finite Size Effects}
\subsection{XY Model}
    So far all our calculations for the XY model were done in the thermodynamic limit ($dk \rightarrow0,\, L\rightarrow\infty$). Now we explore the small system size effects. For that we first define the Fourier transform of correlators as discrete momentum sum which was previously replaced by a momentum integral in Eq.~\eqref{alpheq} and \eqref{Feq},
    \begin{equation}
        \alpha_l(mT)=\frac{1}{\pi L} \sum_{k>0} e^{ilk}\, \langle C_d(mT) \rangle \quad ,\quad F_l(mT)=\frac{i}{\pi L} \sum_{k>0} e^{ilk}\, \langle C_o(mT) \rangle,
    \end{equation}
    where $k=\frac{(2n-1)\pi}{L}, n=1,2,\cdots, L/2$. Fig.~\ref{XY2}(a) implies that for very small system size, steady state concurrence survives for all the drive frequencies and it gets more confined around the special frequencies with increasing system size. One way to intuitively argue this is as follows. The number of multipartite entanglement sectors increases exponentially with system size. Therefore, although 
    for small system size, the local bipartite (pairwise) entanglement can survive, for large systems it gets ``spread out" in exponentially large numbers of multipartite sectors. Therefore, the pairwise entanglement is vanishingly small except at the special frequencies. Consequently critical reset rate ($r_c$) vanishes for all drive frequency in small size limit [Fig.~\ref{XY2}(b)] because we can have long surviving steady state pairwise entanglement in those regime. Vanishing optimal reset rate ($r_m$) for small system [Fig.~\ref{XY2}(c)] implies resetting does not help increasing concurrence in this limit.
  \begin{figure}[ht]
    \centering
    \includegraphics[width=0.32\linewidth]{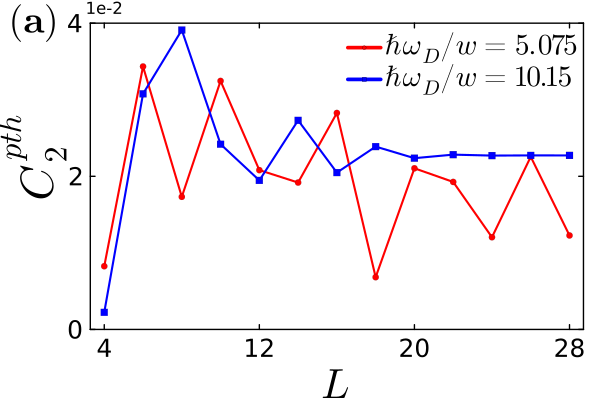}
    \includegraphics[width=0.32\linewidth]{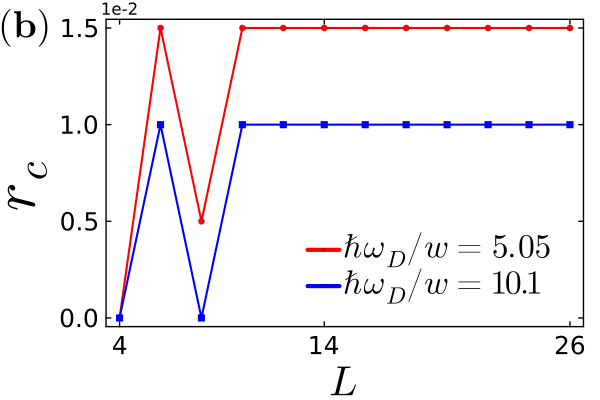}
    \includegraphics[width=0.32\linewidth]{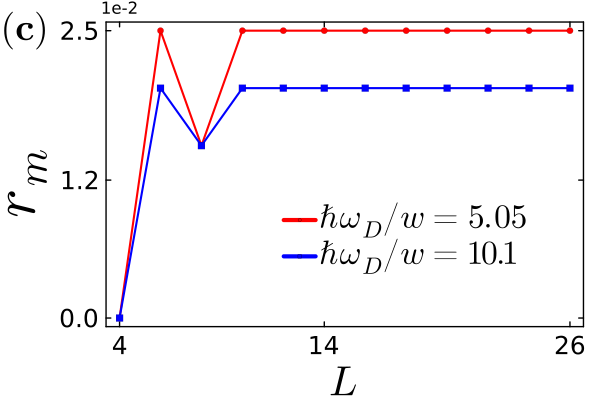}
    \caption{For the PXP model, (a) $C_2^{\rm pth}$ is plotted as a function of chain length($L$) for two slightly shifted special frequencies. (b) $r_c$ and (c) $r_m$  is plotted as a function of chain length for two frequencies (which features both $r_c$ and $r_m$) very close to the slightly shifted special frequencies. We set $\lambda_0/w=10$. See text for more details.}
    \label{PXP2}
\end{figure}
  
\subsection{PXP Model}
   For this model, we have done all the calculations mainly for $L=20$ (main and supplementary text). But this is not of course the highest system size we can achieve for this model. Now we explore system size effect. In Fig.~\ref{PXP2}(a), prethermal averaged concurrence ($C_2^{pth}$) is plotted for two shifted special frequencies as a function of system size ($L$). For higher frequency ($\hbar \omega_D/w=10.15$), the value saturates around $L=20$ whereas for the smaller one ($\hbar \omega_D/w=5.075$) it saturates for higher system size. However the qualitative behavior is similar around $L=20$. Fig.~\ref{PXP2}(b), (c) shows system size dependence of the critical ($r_c$) and optimal reset rate ($r_m$) for two frequencies (close to the special frequencies). Both $r_c$ and $r_m$ saturate quickly around very small system size ($L=10$).

\bibliography{citation}